\documentclass[preprint,12pt,3p]{elsarticle}
\usepackage[utf8]{inputenc}
\usepackage[english]{babel}
\biboptions{sort&compress}

\usepackage{epsfig,amssymb,amsfonts,amsmath,mathtools,bm,color,xcolor,graphicx,braket,adjustbox,esint,upgreek}

\newcommand{\tr}{\mbox{Tr}}
\newcommand{\Hb}{\bar{H}}

\newcommand{\nn}{\nonumber} 
\newcommand{\be}{\begin{equation}}
\newcommand{\ee}{\end{equation}}
\newcommand{\bea}{\begin{eqnarray}}
\newcommand{\eea}{\end{eqnarray}}
\newcommand{\beas}{\begin{eqnarray*}}
\newcommand{\eeas}{\end{eqnarray*}}
\newcommand{\ds}{\displaystyle}

\newcommand{\Br}{\mbox{Br}}

\def\vec#1{\boldsymbol{#1}}

\newcommand{\hSS}{\hat{{\cal S}}_S}
\newcommand{\hSD}{\hat{{\cal S}}_D}

\graphicspath{{./Figs/}}

\journal{Physics Letters B}

\begin{document}

\begin{frontmatter}

\title{Is $Z_{cs}(3982)$ a molecular partner of $Z_c(3900)$ and $Z_c(4020)$ states?}

\author[1,2]{V. Baru}
\author[1]{E. Epelbaum}
\author[1]{A. A. Filin}
\author[3]{C. Hanhart}
\author[4,5]{A. V. Nefediev}

\address[1]{Institut f\"ur Theoretische Physik II, Ruhr-Universit\"at Bochum, D-44780 Bochum, Germany}
\address[2]{Institute for Theoretical and Experimental Physics NRC ``Kurchatov Institute'', Moscow 117218, Russia}
\address[3]{Institute for Advanced Simulation, Institut f\"ur Kernphysik and\\ J\"ulich Center for Hadron Physics, Forschungszentrum J\"ulich, D-52425 J\"ulich, Germany}
\address[4]{P.N. Lebedev Physical Institute of the Russian Academy of Sciences, 119991, Leninskiy Prospect 53, Moscow, Russia}
\address[5]{Moscow Institute of Physics and Technology, 141700, 9 Institutsky lane, Dolgoprudny, Moscow Region, Russia}

\begin{abstract}
We perform an effective-field-theory-based coupled-channel analysis of the recent BES~III data on the $e^+e^-$ annihilation into the final state $K^+(D_s^-D^{*0}+D_s^{*-}D^0)$ in a wide energy range and extract the poles responsible for the formation of the $Z_{cs}(3982)$. We identify two scenarios which provide a similar description of the experimental mass distributions but result in utterly different predictions for the spin partners of the $Z_{cs}(3982)$: although both scenarios are consistent with the $Z_{cs}$ as a $SU(3)$ partner of the $Z_c(3900)$, the $Z_c(4020)$ appears naturally as a spin partner of these states only in one of them (fit 1) while in the other (fit 2) its nature has to be different. Also, the $Z_{cs}(3982)$ has a $J^{P}=1^+$ spin partner near the $\bar D_s^* D^*$ threshold in fit 1, while no such state exists in fit 2. We predict the $\bar{D}_s^*D^*$ invariant mass distribution in the $J^{P}=1^+$ channel
for the reaction $e^+e^-\to K^+D_s^{*-}D^{*0}$ and argue that this line shape can be used to distinguish between the two scenarios once data in this channel are available. 
\end{abstract}


\end{frontmatter}


\section{Introduction}
\label{Sec:intro}

Since 2003 when the famous $\chi_{c1}(3872)$, also known as $X(3872)$,
was discovered by the Belle Collaboration \cite{Belle:2003nnu}, the
number of states in the spectrum of charmonium and bottomonium that do
not fit into the simple quark-model classification scheme grows very
fast. While some of these new hadrons might still be potentially
understood as quark-antiquark mesons, there is a whole class of states
that decay to heavy quarkonia but are charged and for this reason contain at least four quarks. 
Most of them have quantum numbers $I(J^P)=1(1^+)$ and are thus denoted as $Z_Q$, with $Q=c$ or $b$ referring to the heavy $\bar QQ$
content, according to the PDG naming scheme.
Until recently, all known $Z_c$'s and $Z_b$'s, in addition to the $c\bar{c}$ or $b\bar{b}$ pair, respectively, contained only the lightest $u$ or $d$ quarks and, therefore, were subject to the classification according to the $SU(2)$ isospin group. 
In Ref.~\cite{BESIII:2020qkh} the BES~III Collaboration announced a
measurement of the $e^+e^-$ annihilation into the final state
$K^+(D_s^-D^{*0}+D_s^{*-}D^0)$ performed on a data sample with the
integrated luminosity of $3.7$~fb$^{-1}$ in the energy range from
$4.628$ to $4.698$~GeV. An enhancement near the very close by
$D_s^-D^{*0}$ and $D_s^{*-}D^0$ thresholds was observed and fitted with an energy-dependent Breit-Wigner distribution with the parameters $(3982.5^{+1.8}_{-2.6}\pm 2.1)$~MeV and $(12.8_{-4.4}^{+5.3}\pm 3.0)$~MeV for the mass and width, respectively. This state was given the name $Z_{cs}(3982)$. Theoretical predictions of such a state employing different approaches can be found, for example, in Refs.~\cite{Lee:2008uy,Chen:2013wca,Voloshin:2019ilw,Ferretti:2020ewe}.
The experimental observation of this state initiated an active discussion in the community and resulted in a number of theoretical publications devoted to the description of this new state, see, e.g.,
Refs.~\cite{Meng:2020ihj,Wan:2020oxt,Wang:2020kej,Yang:2020nrt,Chen:2020yvq,Cao:2020cfx,Wang:2020htx,Wang:2020iqt,Azizi:2020zyq,Simonov:2020ozp,Ikeno:2020mra,Wang:2020dgr,Guo:2020vmu,Albuquerque:2021tqd,Zhu:2021vtd,Ozdem:2021yvo,Faustov:2021hjs,Chen:2021uou,Ortega:2021enc,Shi:2021jyr,Giron:2021sla,Ding:2021igr,Karliner:2021qok,Wu:2021ezz}. 

In Ref.~\cite{Yang:2020nrt} an analysis of the $\bar{D}_s D^{*0} +
\bar{D}_s^* D^0$ invariant mass spectra was performed in a theory
involving only short-range interactions between the $D$ mesons. The
contact interactions between the $J^{PC}=1^{+-}$ $D^{(*)} \bar D^*$
and $\bar{D}_s D^{*0} + \bar{D}_s^* D^0$ channels were related using
the $SU(3)$ light-quark flavour symmetry. Two mechanisms were
considered for the production: (a) through a triangle $D_{s2} D_s^* D^0$ loop, motivated by the presence of a nearby triangle singularity in the energy range of interest and (b) a point-like production directly feeding two-$D$-meson loops.
The fit was performed in a limited energy range below 4.03~GeV for the
invariant mass of the $\bar{D}_s D^{*0} + \bar{D}_s^* D^0$ system,
that is up to roughly 50 MeV above the threshold, and the exchanges by pseudoscalar Goldstone bosons were neglected. 

The analysis performed in this paper improves the one of
Ref.~\cite{Yang:2020nrt} in several aspects. Most importantly, it
incorporates the whole energy range covered by the BES~III data. This became
possible since we (i) extend the basis of the channels considered to include $\bar{D}_s D^{*0}$, $\bar{D}_s^* D^0$, and $\bar{D}_s^* D^{*0}$ and incorporate coupled-channel effects between them and (ii) find that the triangle diagrams with the $D_{s2}(2573) D_s^* D^{*0}$ and $D_{s1}(2536) D_s^* D^{*0}$ intermediate states also possess triangle singularities within the energy range covered by the BES~III data. We therefore investigate their effect on the line shapes and, in line with Ref.~\cite{BESIII:2020qkh}, find that a tree-level mechanism via $D_{s1} $ is very important for understanding the shape of the spectra at larger invariant masses. 
As a result of the fits to all data, we find two classes of solutions which describe the data equally well but have very different underlying dynamics. Accordingly, the spin partners predicted within the two scenarios by employing heavy quark spin symmetry (HQSS)
are very different, offering a lever arm to distinguish between these scenarios by additional
experiments. For each such scenario we provide a parameter-free prediction for the line shape in the $\bar{D}_s^* D^{*} $ system for $J^{P}=1^+$, which 
is demonstrated to be crucial in this context. It should be noted that also in a broader context, a study of spin symmetry
partners is very valuable, since it not only allows one to disentangle different scenarios within the molecular picture,
but also to distinguish between different interpretations of observed structures~\cite{Cleven:2015era}.

All measurements by BES~III were performed in the energy range close to the state $\psi(4660)$, also known
as $Y(4660)$ ($J^{PC}= 1^{--}$), with a width of $\Gamma = 62^{+9}_{-7}$ MeV. This state is by itself quite interesting
since it apparently does not fit into the spectrum of vector charmonia predicted by various quark models, although
some adjustment of the models might improve the situation --- see the related discussion in Ref.~\cite{Hanhart:2019isz}.
One exotic proposal for this state is a hadrocharmonium structure~\cite{Voloshin:2007dx}, where a compact $\psi(2S)$ core is surrounded by
excited light quarks, which at the same time qualifies as a $f_0(980)\psi(2S)$ hadronic molecule due to the proximity of
the corresponding threshold~\cite{Guo:2008zg}. 
Meanwhile, the state $D_{s2}(2573)$ plays a significant role in the $K(\bar{D}_s D^*+\bar{D}_s^* D)$ invariant mass spectra as is stressed in the experimental analysis of Ref.~\cite{BESIII:2020qkh} as well as in the theoretical study of Ref.~\cite{Yang:2020nrt}. Also, an important role of the $D_{s1}(2536)$ is pointed out in Ref.~\cite{BESIII:2020qkh}.
Naively, this may look surprising because the excitation of the $S$-wave $D_{sJ} \bar{D}_s^{(*)} (J=1,2)$ meson pairs in the $e^+e^-$ annihilation process violates HQSS \cite{Li:2013yka}. Indeed, the light-quark spin-parity states for this meson pair,
$j_L(D_{sJ}) \times j_L(\bar{D}_s^{(*)})= 3/2^+ \times1/2^- = 1^-,
2^-$, see Ref.~\cite{Guo:2014ura} for details, are at odds with the only light-quark spin-parity state $0^+$ allowed in the $e^+e^-$ annihilation through a vector charmonium into a pair of charmed mesons. Thus, a significant violation of HQSS through large couplings of
$Y(4660)$ to the $D_{sJ} \bar{D}_s^{(*)}$ intermediate states is needed for this mechanism to be at work. This, however, would point towards a large $D_{sJ} \bar{D}_s^{(*)}$ molecular component 
of the $Y(4660)$, making it a candidate for a $SU(3)$-flavour partner of $\psi(4230)$ also known as $Y(4230)$\footnote{This state is more widely
known as $Y(4260)$ with the mass label deduced from the discovery experiments. However, more recent measurements
by BES~III called for an adaption of its mass label.}, which is proposed to be a $D_1(2420)\bar D$ molecule in Ref.~\cite{Wang:2013cya}
(this claim is contrasted to the hadrocharmonium picture for this state in Ref.~\cite{Wang:2013kra}; see also the discussion of a possibly large HQSS breaking in production contained in this reference).
This picture is especially intriguing given that the production of the $Z_c(3900)$ in the decay of $Y(4230)$ is claimed to be dominated by triangle 
diagrams\footnote{For a comprehensive review on triangle singularities the interested reader is referred to Ref.~\cite{Guo:2019twa}.}~\cite{Wang:2013cya,Wang:2013hga} in the same way as the triangle loops are important for the $Z_{cs}(3982)$ production in the scenario put forward in Ref.~\cite{Yang:2020nrt} as well as this work. This analogy already suggests a close connection between the strange and the non-strange $Z_c$ states.

Recently, the LHCb Collaboration announced the first observation of exotic states produced in proton-proton collisions and decaying into the
$J/\psi K^+$ final state \cite{LHCb:2021uow}. It remains to be seen whether or not the parameters of the most significant from the observed resonances, the $Z_{cs}(4000)$, can be reconciled with those of the $Z_{cs}(3982)$, indicating that the two structures are manifestations of the same object which reveals itself in both open- and hidden-charm final states. Namely, while the resonance parameters extracted in the experimental analyses were very different, both structures were successfully described in Ref.~\cite{Ortega:2021enc} within the same coupled-channel model using the formalism of the chiral constituent quark model. A similar claim that the LHCb and BES~III structures may be siblings is contained in Ref.~\cite{Yang:2020nrt}. In this paper we focus on the BES~III data and postpone an analysis of the LHCb data to a later, more complete, study.

\section{Formalism}

\subsection{Contact interactions}
\label{Sec:CT}

We start from a short introduction to the light-quark $SU(3)$ flavour group. Consider a generic state of the form $|M_A\bar{M}^{\prime B}\rangle$, where 
$M$ and $M'$ are two heavy-light mesons which can be either pseudoscalar ($P=D,D_s$) or vector ($V=D^*,D_s^*$)
states, so that all possible combinations of the form
$D^{(*)}_{(s)}\bar{D}^{(*)}_{(s)}$ are captured this way. Here,
$A,B=1,2,3$ refer to the $SU(3)$ indices. The state
$|M_A\bar{M}^{\prime B}\rangle$ belongs to a reducible representation $3\otimes\bar{3}$ and can be decomposed into a sum of irreducible representations ($3\otimes\bar{3}=1\oplus 8$) as
\be
|M_A\bar{M}^{\prime B}\rangle=\frac{1}{\sqrt{3}}\delta_A^B|M\bar{M}',S\rangle+\frac{1}{\sqrt{2}}\sum_{i=1}^8\left(\lambda_i\right)_A^B|M\bar{M}',O;i\rangle,
\ee
where $\{\lambda_i\}$ ($i=1,\ldots,8$) are the Gell-Mann matrices and we
introduced the mutually orthogonal and normalised to unity singlet ($S$) and octet ($O$) states,
\be
|M\bar{M}',S\rangle=\frac{1}{\sqrt{3}}|M_A\bar{M}^{\prime A}\rangle,\quad
|M\bar{M}',O;i\rangle=\frac{1}{\sqrt{2}}\left(\lambda_i\right)_B^A|M_A\bar{M}^{\prime B}\rangle,
\quad i=1,\ldots,8.
\ee

If the interaction $\hat{V}$ is $SU(3)$ symmetric, then its
non-vanishing matrix elements in the flavour space can be parameterised via only two potentials as 
\be
\langle M\bar{M}',S|\hat{V}|M\bar{M}',S\rangle=V_S,\quad
\langle M\bar{M}',O;i|\hat{V}|M\bar{M}',O;j\rangle=V_O\delta_{ij},
\label{hatV}
\ee
so that it is straightforward to find that (no summation over $A$ and $B$ is implied here) 
\be
\langle M_A\bar{M}^{\prime A}|\hat{V}|M_B\bar{M}^{\prime B}\rangle=
\left\{
\begin{tabular}{lll}
$\frac13(V_S+2V_O)$,&~~& $A=B$,\\
$\frac13(V_S-V_O)$,&~~& $A\neq B$
\end{tabular}
\right.
\label{matrelem1}
\ee
and
\be
\langle M_A\bar{M}^{\prime B}|\hat{V}|M_A\bar{M}^{\prime B}\rangle=V_O,\quad A\neq B.
\label{matrelem2}
\ee

It proves convenient to redefine the interactions such that $\{V_S,V_O\}\to\{V_s,V_t\}$, where $V_s$ and $V_t$ correspond to the isoscalar and isotriplet potentials, respectively. To this end we decompose an arbitrary $SU(2)$ state $|M_a\bar{M}^{\prime b}\rangle$ ($a,b=1,2$) as 
\be
|M_a\bar{M}^{\prime b}\rangle=\frac{1}{\sqrt{2}}\delta_a^b|M\bar{M}',[0]\rangle+\frac{1}{\sqrt{2}}\sum_{i=1}^3\left(\tau_i\right)_a^b|M\bar{M}',[1];i\rangle,
\ee
where $\{\tau_i\}$ ($i=1,2,3$) are Pauli matrices and the total isospin of the state is quoted in square brackets, to find 
\bea
\langle M\bar{M}',[0]|\hat{V}|M\bar{M}',[0]\rangle&=&\frac13(2V_S+V_O)\equiv V_s,\nonumber\\[-2mm]
\label{su3pot1}\\[-2mm]
\langle M\bar{M}',[1]|\hat{V}|M\bar{M}',[1]\rangle&=&V_O\equiv V_t.\nonumber
\eea

Then, it is straightforward to evaluate the potential for the state $|M\bar{M}',[1/2]\rangle$, which is the central object of this study:
\be
\langle M\bar{M}',[1/2]|\hat{V}|M\bar{M}',[1/2]\rangle=\langle M\bar{M}',[1]|\hat{V}|M\bar{M}',[1]\rangle=V_O=V_t.
\label{su3pot2}
\ee
This result coincides with the findings of
Refs.~\cite{HidalgoDuque:2012pq,Yang:2020nrt}. Therefore, in order to
build the potential in the $I=1/2$ sector it is sufficient to derive
it in the isovector channel. To this end we consider the HQSS Lagrangian for the $D$ mesons in the isovector channel which, at leading order ${\cal O}(Q^0)$, reads \cite{Mehen:2011yh}
\bea
{\cal L}^{(0)}_{HH}&=&{\rm Tr}\left[H^\dagger_a \left(i \partial_0 +\frac{\vec\nabla^2}{2 \bar{M}}\right) H_a\right]
+{\rm Tr}\left[\Hb^\dagger_a\left(i\partial_0+\frac{\vec\nabla^2}{2\bar{M}}\right)\Hb_a\right] \nn \\ 
&+&\frac{\delta}{4}{\rm Tr}[H^\dagger_a\sigma^i H_a \sigma^i]
+\frac{\delta}{4}{\rm Tr}[\Hb^\dagger_a\sigma^i\Hb_a \sigma^i]\label{Lag0}\\
&-&\frac{C_{10}}{8}{\rm Tr}[\bar{H}^\dagger_a\tau^A_{aa^\prime}H^\dagger_{a^\prime}H_b\tau_{bb^\prime}^A\bar{H}_{b^\prime}] 
-\frac{C_{11}}{8}{\rm Tr}[\bar{H}^\dagger_a\tau^A_{aa^\prime}\sigma^i H^\dagger_{a^\prime}H_b\tau_{bb^\prime}^A\sigma^i\bar{H}_{b^\prime}],\nn
\eea
where $\sigma$ and $\tau$ denote the spin and isospin Pauli matrices, respectively, and the trace is taken in the spin space. As before, $a$ and $b$ are the isospin indices, and the isospin matrices are normalised as $\tau^A_{ab} \tau^B_{ba} =2\delta^{AB}$. The mass $\bar{M}$ in the kinetic terms is the spin-averaged $D$ meson mass, 
$\bar{M}=(3m_*+m)/4$, and $\delta=m_*-m\approx 140$~MeV. The terms in the first line in Eq.~\eqref{Lag0} stand for the leading heavy and anti-heavy meson Lagrangian of Refs.~\cite{Wise:1992hn,Burdman:1992gh,Yan:1992gz},
written in the two-component notation of Ref.~\cite{Hu:2005gf}. 
The terms proportional to the low-energy constants (LECs) $C_{10}$ and $C_{11}$ correspond to the ${\cal O}(Q^0)$ $S$-wave contact interactions~\cite{Mehen:2011yh,AlFiky:2005jd}.
The superfields $H_a$ and $\bar{H}_a$ of the heavy-light mesons read
\be
H_a=P_a+V^i_a\sigma^i,\quad\bar{H}_a=\bar{P}_a-\bar{V}^i_a\sigma^i,
\label{Hs}
\ee 
where $P=D$ and $V=D^*$ for mesons and, similarly, $\bar{P}=\bar{D}$ and $\bar{V}=\bar{D}^*$ for antimesons. 

Hereinafter we refer to open-charm channels as elastic. Then, in the $J^P=1^+$ channel
with the basis elastic states\footnote{Here, the individual partial waves are labelled as $^{2S+1}L_J$ with $S$, $L$, and $J$ denoting the total
spin, orbital angular momentum and the total angular momentum of the two-meson system, respectively.} 
\be
1^+:\hspace*{0.2cm}\{\bar{P}V(^3S_1),P\bar{V}(^3S_1),V\bar{V}(^3S_1)\},
\label{1p}
\ee
the HQSS-respecting $S$-wave order-$Q^0$ contact potential reads
\be
V^{\rm CT}[1^+]=\left(
\begin{array}{cccccc}
\mathcal{C}_d+\frac12\mathcal{C}_f \hspace{0.3cm} & \frac12\mathcal{C}_f & -\frac1{\sqrt2}\mathcal{C}_f \\
\frac12\mathcal{C}_f \hspace{0.3cm} & \mathcal{C}_d+\frac12\mathcal{C}_f \hspace{0.3cm} & \frac1{\sqrt2}\mathcal{C}_f \\
-\frac1{\sqrt2}\mathcal{C}_f\hspace{0.3cm} & \frac1{\sqrt2}\mathcal{C}_f & \mathcal{C}_d
\end{array}\right),
\label{v3x3}
\ee
where two linear combinations of the LECs from the Lagrangian \eqref{Lag0} were introduced as 
$\mathcal{C}_d=\frac18(C_{11}+C_{10})$ and
$\mathcal{C}_f=\frac18(C_{11}-C_{10})$, with the subscript $d$ ($f$)
referring to the diagonal (off-diagonal) terms. 

The basis vectors for $J^{PC}$ states in various elastic channels with a given $C$-parity read 
\begin{eqnarray}
&&1^{+-}:\hspace*{0.2cm}\{P\bar{V}(^3S_1,-),V\bar{V}(^3S_1)\},\nonumber\\
&&1^{++}:\hspace*{0.2cm}\{P\bar{V}(^3S_1,+) \},\label{basisvec}\\
&&2^{++}:\hspace*{0.2cm}\{V\bar{V}(^5S_2)\},\nonumber
\end{eqnarray}
where the sign in parentheses gives the $C$-parity of the corresponding $P\bar{V}$ states defined as 
\be
\ket{P\bar{V},\pm}=\frac{1}{\sqrt{2}}(\ket{P\bar{V}}\pm\ket{\bar{P}V}),
\label{Eq:Cpar}
\ee
with the $C$-parity transformation defined as ${\hat C}M=\bar{M}$ for any meson $M$. Then it is easy to find that the leading-order contact potentials in the elastic channels read
\be
V^{\rm CT}[1^{+-}]=
\left(\begin{array}{cccc}
\mathcal{C}_d \hspace{0.3cm} & \mathcal{C}_f \\
\mathcal{C}_f \hspace{0.3cm} & \mathcal{C}_d 
\end{array}\right),\quad
V^{\rm CT}[1^{++}]=V^{\rm CT}[2^{++}]={\cal C}_d+{\cal C}_f.
\label{vfull}
\ee
In particular, we see that once the poles in the $1^{+-}$ sector are settled to provide a decent description of the line shapes, those in the $1^{++}$ and $2^{++}$ channels are fixed and driven by identical potentials~\cite{Nieves:2012tt,Baru:2016iwj}.
The latter potentials were previously used to a success in studies of the isoscalar states of charmonium (see, e.g., a review article~\cite{Guo:2017jvc} and references therein) and isovector states of bottomonium \cite{Wang:2018jlv,Baru:2019xnh,Baru:2020ywb}.

According to Eq.~(\ref{su3pot2}), the effective potential of Eq.~\eqref{v3x3} will be used in what follows for the three-channel problem 
$\{D_s^-D^{*0},D^{*-}_s D^0,D^{*-}_sD^{*0}\}\coloneqq \{1,2,3\}$, where the thresholds are ordered from the lowest to the highest in energy. 
For a given set $J^{P(C)}$ the system of the partial-wave-decomposed coupled-channel Lippmann-Schwinger equations reads
\be
T_{\alpha\beta}(\sqrt{s},p,p')=V^{\rm eff}_{\alpha\beta}(p,p')-\sum_\gamma \int \frac{d^3q}{(2\pi)^3} V^{\rm eff}_{\alpha\gamma}(p,q){G_\gamma(\sqrt{s},q)}T_{\gamma\beta}(\sqrt{s},q,p'),
\label{Eq:JPC}
\ee
where $\alpha$, $\beta$ and $\gamma$ label the basis vectors defined in Eqs.~(\ref{1p}) and (\ref{basisvec}), the effective potential is defined in Eqs.~\eqref{v3x3} and \eqref{vfull},
and the scattering amplitude $T_{\alpha\beta}$ is related to the invariant amplitude ${\cal M}_{\alpha\beta}$ as
\bea
T_{\alpha\beta}=-\frac{{\cal M_{\alpha\beta}}}{\sqrt{(2m_{1,\alpha})(2m_{2,\alpha})(2m_{1,\beta})(2m_{2,\beta})}},
\eea
with $m_{1,\alpha}$ and $m_{2,\alpha}$ ($m_{1,\beta}$ and $m_{2,\beta}$) being the masses of the $D_{(s)}^{(*)}$ mesons in the channel $\alpha$ ($\beta$).
The two-body propagator for the given set $J^{PC}$ takes the form 
\bea
G_\gamma=\left(q^2/(2\mu_\gamma)+m_{1,\gamma}+m_{2,\gamma}-\sqrt{s}-i\epsilon\right)^{-1},
\eea
where the reduced mass is
\be
\mu_\gamma=\frac{m_{1,\gamma}m_{2,\gamma}}{m_{1,\gamma}+m_{2,\gamma}}
\ee
and $\sqrt{s}$ is the total energy of the system.

Finally, to render the loop integrals well defined, we introduce a sharp ultraviolet cutoff $\Lambda$ which needs to be larger than all
typical three-momenta related to the coupled-channel dynamics. For the results presented below we choose $\Lambda=1$~GeV, but we have also
verified that our results are almost cutoff independent by repeating the fit with cutoffs varied over a sufficiently wide range from 0.8 to 1.3 GeV.

The extension of the approach to incorporate the $\eta$-meson exchange, as the only representative of the $SU(3)$ Goldstone boson octet allowed here, 
is delegated to future studies. This requires an inclusion of the tensor interactions and their appropriate renormalisation along the line of the recent studies in the context of the $Z_b(10610)$ and $Z_b(10650)$ \cite{Wang:2018jlv,Baru:2019xnh} and the LHCb pentaquarks \cite{Du:2021fmf}.

\subsection{Production amplitude}

\begin{table}[t]
\begin{center}
\begin{tabular}{|c|c|c|c|}
\hline 
$\,$ & $D_{s2}D_s^*D^0$& $D_{s2}D_s^*D^{*0}$ & $D_{s1}D_s^*D^{*0}$ \\
\hline 
$\sqrt{s}$ [GeV]& [4.6813, 4.7117] & [4.6813, 4.6894] & [4.6474, 4.6513]\\
\hline
$m_{23}$ [GeV]& [3.977, 4.003] & [4.1191, 4.1262] & [4.1191, 4.1226]\\
\hline
\end{tabular}
\end{center}
\caption{The ranges of $\sqrt{s}$ and the invariant mass of the final $D$ mesons ($m_{23}$), where the triangle singularity occurs for various intermediate states involving the $D_{s2}$ and $D_{s1}$ excited mesons, if they had no widths. All energies are within the range measured by BES~III. The minimal value of $\sqrt{s}$ corresponds to the maximum of $m_{23}$ and vice versa.}\label{tableII_triangle_range}
\end{table}

\begin{figure}[t]
\centerline{\includegraphics[width=0.9\textwidth]{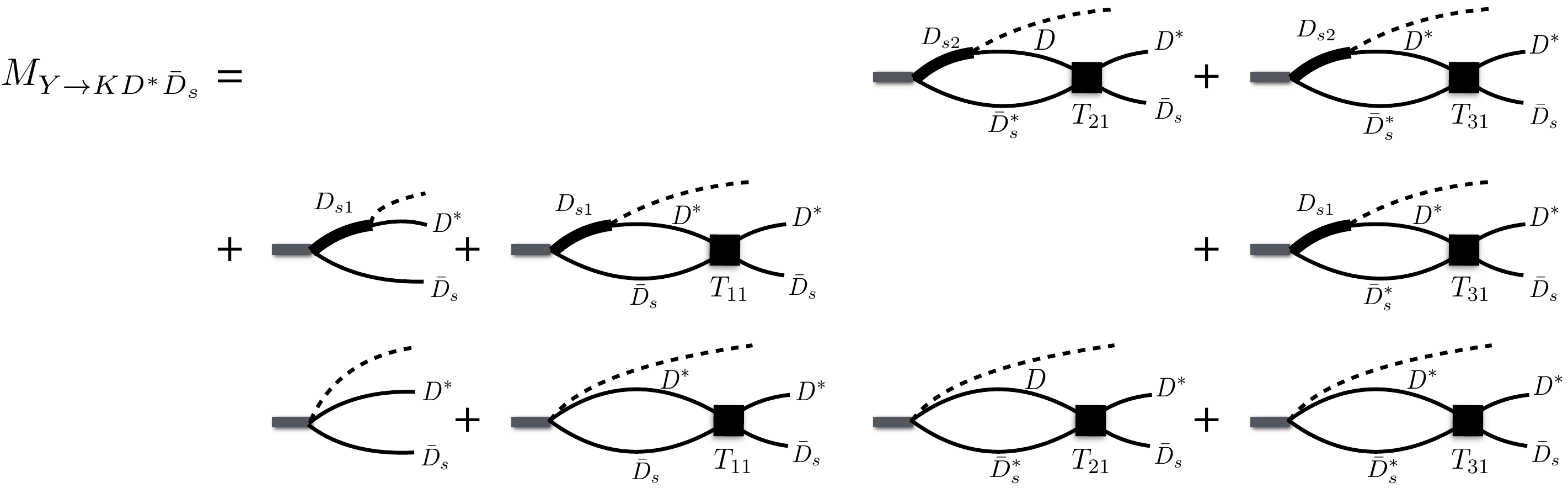}}\vspace*{0.4cm}
\centerline{\includegraphics[width=0.9\textwidth]{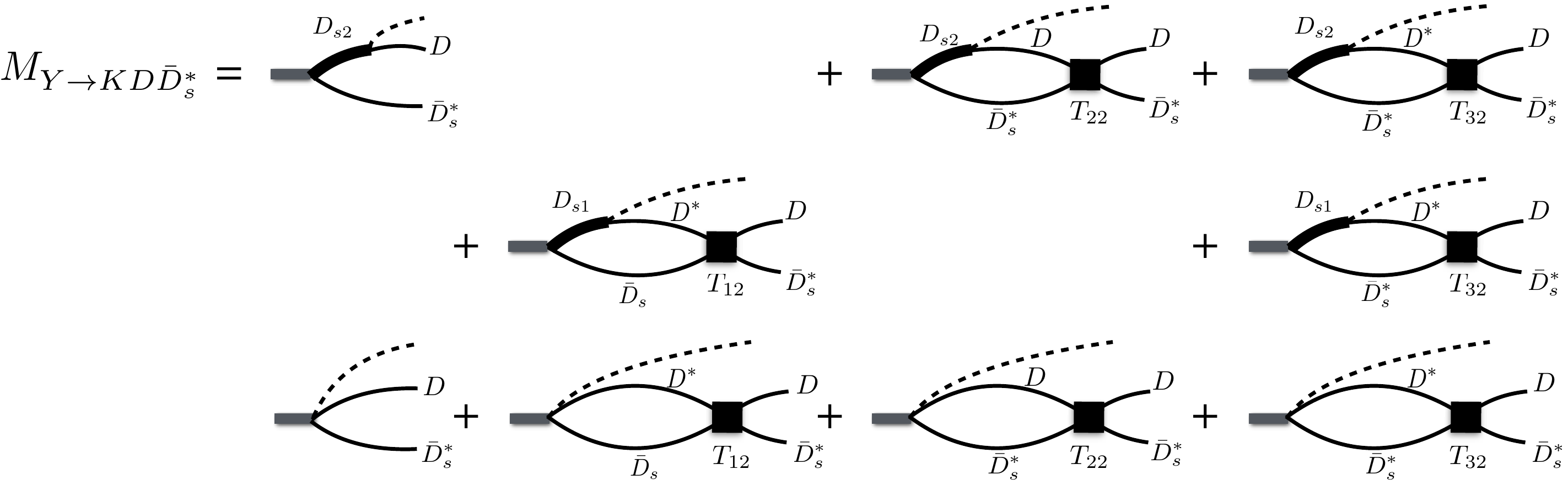}}
\caption{Diagrams contributing to the transition amplitude for the $K\bar{D}_s D^*$ and $K\bar{D}_s^* D$ final states. The first and second rows in each channel contain the contributions which proceed via the excitation of the $D_{s2}(2573)$ and $D_{s1}(2536)$ resonance states, respectively, the third row shows the diagrams for the point-like production mechanism via the $D$-meson loops. The dashed line denotes the kaon and the bold grey line is for the Y(4660). All $D$ mesons are explicitly tagged in the diagrams. \label{fig:diag} }
\end{figure}

It is a natural assumption in the molecular approach that a point-like production mechanism generates first a pair of particles in the elastic channels, which then interact strongly in the final state to produce visible structures in the line shapes near the elastic thresholds. For example, such a production mechanism in the molecular picture provides a very good description of the line shapes for the $Z_b(10610)/Z_b(10650)$ states measured by Belle \cite{Wang:2018jlv,Baru:2019xnh,Baru:2020ywb} and the pentaquark states discovered by LHCb \cite{Du:2019pij,Du:2021fmf}. 
However, since it is argued in Ref.~\cite{BESIII:2020qkh} that the
states $D_{s2}(2573)$ and $D_{s1}(2536)$ provide the most important
contributions to the experimental background in the observed $K(\bar{D}_s
D^*+\bar{D}_s^* D)$ invariant mass spectra, their
contributions need to be considered as they might encode important information
about the nature of the $Y(4660)$ as discussed in the introduction.
Specifically, since the $D_{s1}\bar{D}_s^*$ and $D_{s2}\bar{D}_s^*$
thresholds at 4.647 GeV and 4.681 GeV, respectively, are within the
energy range covered by the BES~III measurement, the amplitude which
proceeds via these intermediate states is enhanced. In addition, the
triangle singularities generated by the $D_{s2}D_s^* D$, $D_{s2}D_s^*
D^*$, and $D_{s1}D_s^* D^*$ loops are not far from the physical region
even if the widths of the $D_{sJ}$ mesons are included --- see
Table~\ref{tableII_triangle_range} for details.

Therefore, the properties of the $Z_{cs}$ and its spin partners can be extracted from the current data assuming HQSS only if 
an interplay of different production mechanisms is explored in the whole energy range covered by the BES~III measurement --- see also the discussion in Ref.~\cite{Du:2020vwb}. 
The diagrams considered in the current study are shown in Fig.~\ref{fig:diag}. 
In order to minimise the number of parameters we follow the approach
of Ref.~\cite{Yang:2020nrt} and assume that the production proceeds
solely via $Y(4660)$ with the $S$-wave couplings of $Y(4660)$ to
$D_{sJ} \bar{D}_s^{(*)}$ being energy-independent constants. This is
justified since all measurements by BES~III were performed in the
energy range close to this state. Then, the parity and the total angular momentum conservation ensure that the $Y(4660)$ can decay into the $D_{s1}(2536) \bar{D}_s^{(*)}$ and $ D_{s2}(2573) \bar{D}_s^*$ meson pairs, with subsequent decays of the $D_{s1}$ to $D^* K$ and $D_{s2}$ to $D^{(*)} K$, respectively. This is encoded in the diagrams shown in the first two rows for each of the two final states $K \bar{D}_sD^*$ and $K \bar{D}_s^* D$. 

\begin{figure}[t]
\centering
\includegraphics[width=0.49\textwidth]{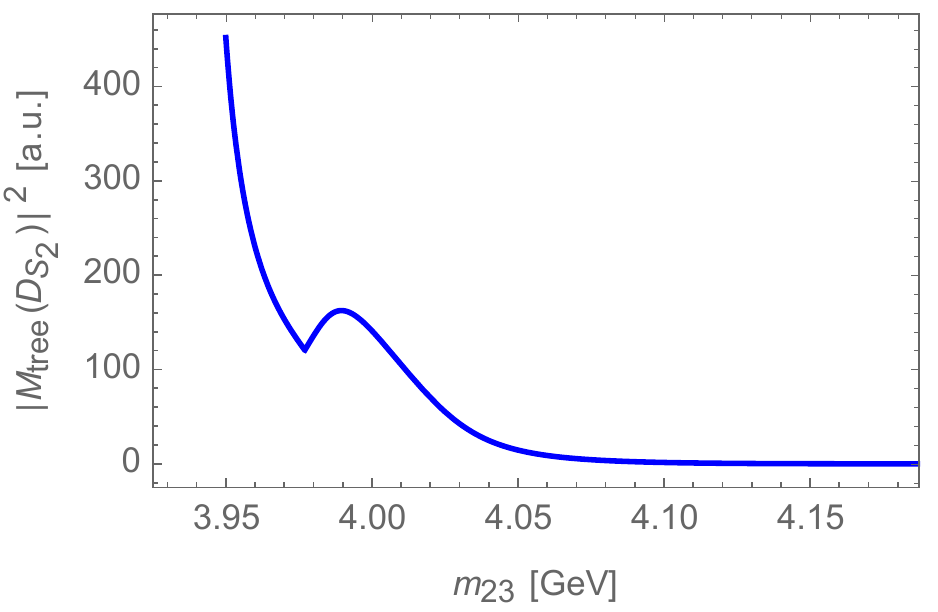}
\includegraphics[width=0.49\textwidth]{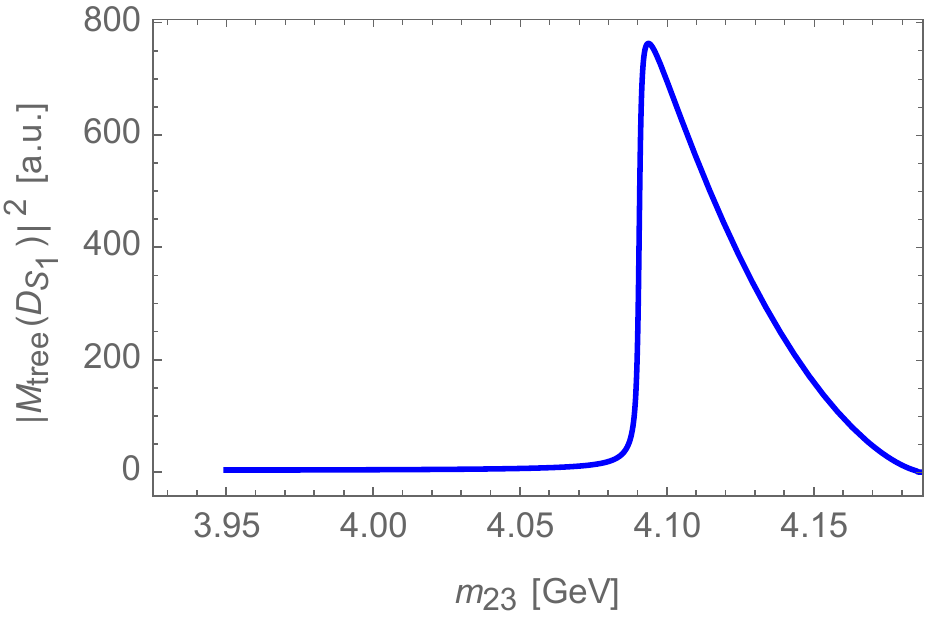}
\caption{The contributions of the tree-level diagrams in $D$ waves (see Eqs.~\eqref{eq:A1} and\eqref{eq:MD2}) with the $D_{s2}(2573)$ (left panel) and $D_{s1}(2536)$ (right panel)  
versus the invariant mass of the $\bar{D}_s D^*/ \bar{D}_s^* D$ meson pairs ($m_{23}$) at $\sqrt{s}=4.681$ GeV.}
\label{fig:tree}
\end{figure}

\begin{figure}[t]
\centering
\includegraphics[width=0.49\textwidth]{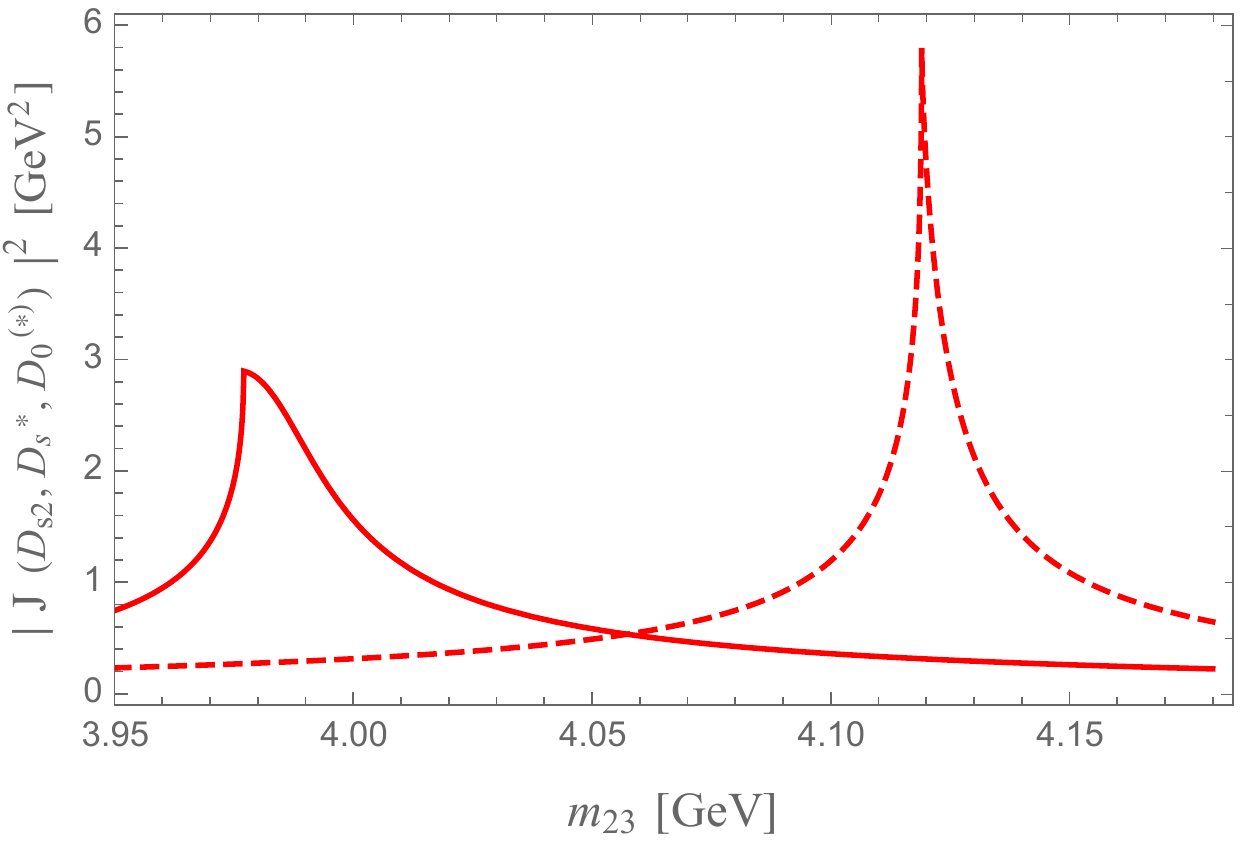}
\includegraphics[width=0.49\textwidth]{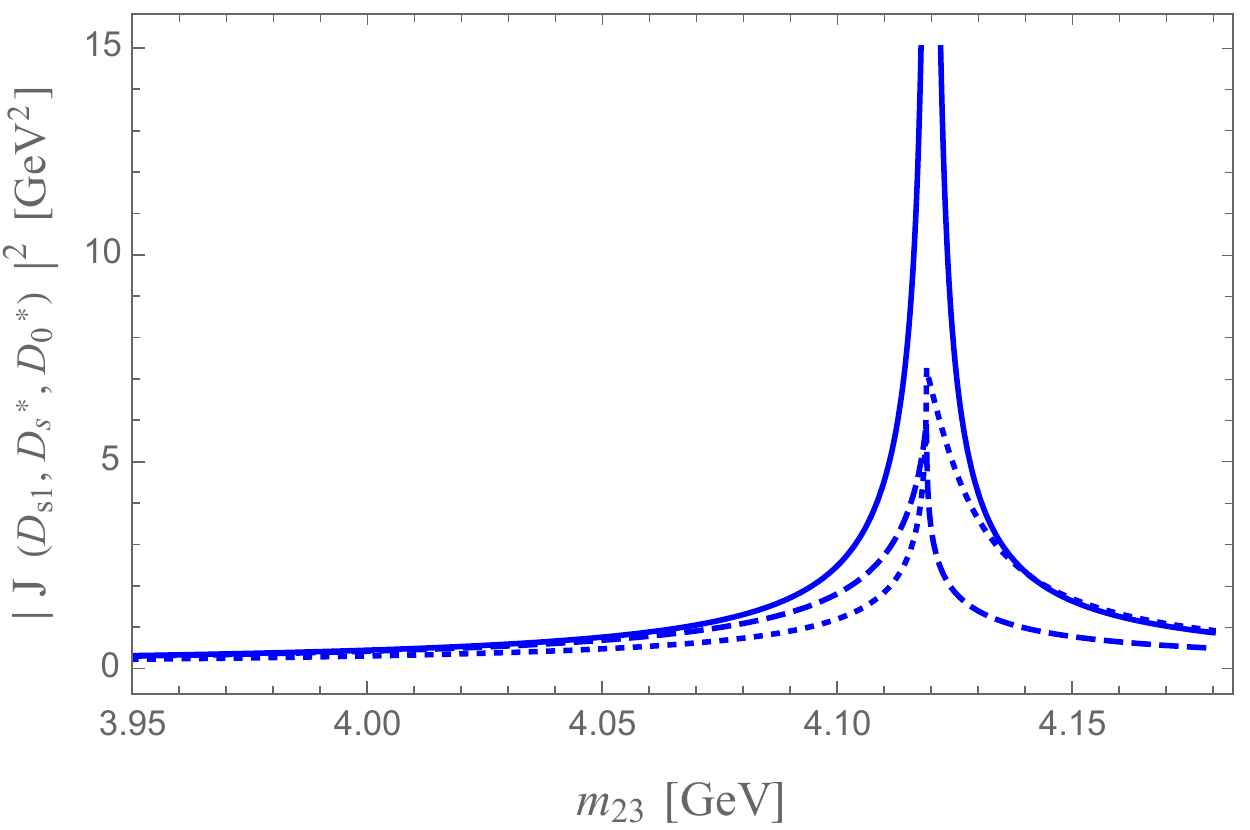}
\caption{ The triangle functions defined by Eq.~\eqref{Eq:J} versus the invariant mass of the $\bar{D}_s D^*/ \bar{D}_s^*D$ meson pairs. 
Left panel: the triangle loops for the intermediate $D_{s2}D_s^*D$ (solid red) and $D_{s2}D_s^*D^*$ (dashed red) states at $\sqrt{s}=4.681$ GeV. Right panel: the triangle loops for the intermediate state $D_{s1}D_s^* D^*$ at $\sqrt{s}=4.648$ GeV (solid blue), where it is in the regime of the singularity (see Table~\ref{tableII_triangle_range}), and at the energies measured by BES~III, namely $\sqrt{s}=4.641$ GeV (dotted blue) and $\sqrt{s}=4.661$ GeV (dashed blue). The physical widths of the $D_{s2}(2573)$ and $D_{s1}(2536)$ are included \cite{Zyla:2020zbs}.}
\label{fig:triangles}
\end{figure}

The tree-level diagrams with the excited $D_{s2}$ and $D_{s1}$ lead to pronounced enhancements in the invariant mass distribution of the two $D$-mesons --- see Fig.~\ref{fig:tree} for an illustration. The intermediate states in these diagrams, that is $D_{s2} \bar{D}_s^*$ and $D_{s1} \bar{D}_s$, respectively, can go on shell. The pole from the $D_{s2}$ emerges roughly 200 MeV above
the $KD$ threshold, which leads to an enhanced production of the $KD$ state in the corresponding energy range. Then, as a cross-channel effect, the invariant mass spectrum in the channel $\bar{D}_s D^*/ \bar{D}_s^* D$ shows a very pronounced structure near the $\bar{D}_s D^*/ \bar{D}_s^* D$ threshold. The effect of the $D_{s1}$ exchange demonstrates quite an opposite pattern: since the $D_{s1}$ mass is only about 30 MeV above the $KD^*$ threshold, the peak in the two $D$-meson invariant mass spectrum shows up at larger invariant masses. In addition to the structures from the tree-level diagrams, the triangle singularities from the $D_{s2} D_s^* D$, $D_{s2} D_s^* D^*$, and $D_{s1} D_s^* D^*$ loops may reveal themselves in the energy region covered by the BES~III data, as illustrated in Fig.~\ref{fig:triangles}. 

The spin-angular structure of the tree-level and triangle
contributions from Fig.~\ref{fig:diag} can be derived from the
effective Lagrangian that relates the $S$-wave charmonia with the $S$-wave $D_{sJ}\bar{D}^{(*)}$-meson pairs \cite{Guo:2014ura},
\be 
{\cal L}_{\rm JT\bar H}=g_{\rm JT H}\tr[J\, {T^i_a}^\dagger \bar H_a^\dagger \sigma^i] +{\rm h.c.},
\label{L_YTH}
\ee
and those for the subsequent decay of the $D_{sJ}$-meson into $D^{(*)} K$ in an $S$ wave \cite{Guo:2020oqk},
\be
{\cal L}_{\rm THK}^{(S)}=ig_{\rm THK}^{(S)}(\vec{D}_{s1\;a}\cdot\vec{D}^{*\dagger}_b)\partial_0 \phi_{ab}^\dagger+{\rm h.c.},
\label{L_THKS}
\ee
and $D$ wave \cite{Colangelo:2005gb}, 
\be
{\cal L}_{\rm THK}^{(D)}=g_{\rm THK}^{(D)}\tr[{T^i_a} \sigma^j H_b^\dagger ] \partial_i \partial_j\phi_{ab}^\dagger+{\rm h.c.}.
\label{L_THKD}
\ee
Here, $J=\vec{Y}\cdot\boldsymbol{\sigma}+\eta_c$ is the spin multiplet for the $S$-wave charmonia, 
\be
T_a^i=P_{2a}^{ij}\sigma^j +\displaystyle\sqrt\frac23 P_{1a}^{i} + i \sqrt\frac16 \epsilon_{ijk}P_{1a}^{j}\sigma^k
\label{T_DsJ}
\ee
is the spin multiplet of the heavy-light mesons with the total angular momentum of the light quarks $j_L^P=3/2^+$ (the fields $P_{1a}$ and $P_{2a}$ annihilate the charmed mesons $D_{s1}(2536)$ and $D_{s2}(2573)$, respectively), the spin multiplets of the $j_L^P=1/2^-$ heavy-light mesons are defined in Eq.~(\ref{Hs}) above, and $\phi$ is the pseudoscalar field.

In addition, the effective Lagrangian connecting the $S$-wave charmonia with the $S$-wave $D$-meson pairs and a pseudoscalar field reads \cite{Mehen:2011yh} 
\bea 
{\cal L}_{\rm HHJ \phi}= 
\frac{1}{4}g_{\rm HHJ\phi}\tr[J \bar H_b^\dagger H_a^\dagger ]{u^0_{ab}}^\dagger +{\rm h.c.},
\label{L_YHHK}
\eea
where the appearance of the zeroth component of $u$ (time derivative in the formula below which translates into the energy) is a consequence of chiral symmetry. The pseudoscalar Goldstone bosons are parameterised as 
\bea
&u_0=i\left(u^\dagger \dot{u}-u\dot{u}^\dagger\right),\quad \ds u = \exp \left( \frac{i\phi}{\sqrt{2}f}\right),& \\
&\ds \phi = \begin{pmatrix}
 {\frac{1}{\sqrt{2}}\pi ^0 +\frac{1}{\sqrt{6}}\eta _8 } & {\pi^+ } & {K^+ } \\
 {\pi^- } & {-\frac{1}{\sqrt{2}}\pi ^0 +\frac{1}{\sqrt{6}}\eta _8} & {K^0 } \\
 { K^-} & {\bar{K}^0 } & {-\frac{2}{\sqrt{6}}\eta_8 } \\
 \end{pmatrix},&\nonumber 
 \label{eq:u-phi-def}
 \eea 
where $f$ is the $SU(3)$ Goldstone bosons decay constant; $f=f_{\pi}=92.2$ MeV \cite{Zyla:2020zbs}.

The sum of the amplitudes corresponding to the diagrams shown in Fig.~\ref{fig:diag} for the final states $K^+ D^{*0} {D}_s^{-}$ and $K^+ D^{0} {D}_s^{*-}$ reads
\be
M[\alpha]=-\sqrt{2m_Y}\sum_{l=0,2}{\cal S}_l[\alpha]M_l[\alpha] k^l=-\sqrt{2m_Y} (\hSD[\alpha]M_D[\alpha]k^2+\hSS[\alpha]M_S[\alpha]), 
\ee
with $m_Y$ for the mass of the $Y(4660)$.
Here, $\alpha=1$ and 2 correspond to the $\bar{D}_sD^*$ and
$\bar{D}_s^*D$ states, respectively, the appropriately normalised
spin-orbit structures are $\hSS[\alpha]=
\vec{\varepsilon}_{Y}\cdot\vec{\varepsilon}[\alpha]$ and
$\hSD[\alpha]=-\frac{3}{\sqrt{2}}
\varepsilon_{Y_i}\varepsilon_j[\alpha] n_{ij} $, where $n_{ij} = n_i
n_j-\delta_{ij}/3$ with $n_i=\vec{k}_i/k $, and the magnitude of the kaon
momentum in the rest frame of the decaying particle is $k =\lambda^{1/2}(s,m_1^2,m_{23}^2)/(2\sqrt{s})$ with $m_{23}$ being the invariant mass of the $D$-meson pair. 
The importance of the $S$-wave kaon production amplitude was stressed in Ref.~\cite{Du:2020vwb}. Finally, for the $D$- and $S$-wave amplitudes $M_D[\alpha]$ (see the first two rows for each final state in Fig.~\ref{fig:diag}) and $M_S[\alpha]$ (see the last row for each final state in Fig.~\ref{fig:diag}) with $\alpha=1,2$ one has 
\bea
M_D[1]&=&g_{D_{s2}}\left[J(D_{s2},D_s^*,D)T_{21}+\frac{3\sqrt2}{4}J(D_{s2},D_s^*,D^*)T_{31}\right]\label{eq:A1}\\
 &+&g_{D_{s1}}\left[\frac{2m_{D_{s1}}}{t[1]-m_{D_{s1}}^2+i\Gamma_{D_{s1}}m_{D_{s1}}}+J(D_{s1},D_s,D^*)T_{11}-\frac{\sqrt{2}}{4}J(D_{s1},D_s^*,D^*)T_{31}\right]\nonumber\\
M_S[1]&=&\lambda g_{D_{s1}} \omega(k)\left[\frac{2m_{D_{s1}}}{t[1]-m_{D_{s1}}^2+i\Gamma_{D_{s1}}m_{D_{s1}}}+J(D_{s1},D_s,D^*)T_{11}-\frac{\sqrt{2}} 
{4}J(D_{s1},D_s^*,D^*)T_{31}\right]\nonumber\\
 &+&g\,\omega(k) \left[-1+ J_0(D_s,D^*)T_{11}-J_0(D_s^*,D)T_{21}+\sqrt{2}J_0(D_s^*,D^*)T_{31}\right],\\ \label{eq:MD2}
M_D[2]&=&g_{D_{s2}}\left[\frac{2m_{D_{s2}}}{t[2]-m_{D_{s2}}^2+i\Gamma_{D_{s2}}m_{D_{s2}}}+J(D_{s2},D_s^*,D)T_{22}+\frac{3\sqrt2}{4} 
J(D_{s2},D_s^*,D^*)T_{32}\right]\nonumber\\
 &+&g_{D_{s1}}\left[J(D_{s1},D_s,D^*)T_{12}-\frac{\sqrt{2}}{4}J(D_{s1},D_s^*,D^*)T_{32}\right],\\
M_S[2]&=&\lambda g_{D_{s1}} \omega(k)\left[J(D_{s1},D_s,D^*)T_{12}-\frac{\sqrt{2}}{4}J(D_{s1},D_s^*,D^*)T_{32}\right]\label{eq:A2}\\
 &+&g\,\omega(k)\left[1-J_0(D_s^*,D)T_{22}+J_0(D_s,D^*)T_{12}+\sqrt{2}J_0(D_s^*,D^*)T_{32}\right],\nonumber
\eea
where $t[1]=m^2(KD^*)$ and $t[2]=m^2(KD)$, with $m^2(XY)$ referring to the invariant mass of the $XY$ system, and the kaon energy is $\omega(k)=\sqrt{k^2+m_K^2}$. The relative coefficients in the brackets are fixed by HQSS.

The integrals $J(a,b,c)$ and $J_0(b,c)$ are the triangle and scalar loop functions defined as
\bea
J(a,b,c)&=&\int\frac{d^4 l}{(2\pi)^4}\frac{8m_a m_b m_c i}{[(l+p_{ab})^2-m_a^2+i\Gamma_a m_a][l^2-m_b^2+i\epsilon][(l-p_{bc})^2-m_c^2+i\epsilon]},\label{Eq:J}\\
J_0(b,c)&=&\int\frac{d^3 l}{(2\pi)^3}\frac{\theta{(l-\Lambda)}}{l^2/(2\mu_{bc})+m_{b}+m_{c}-\sqrt{s}-i\epsilon},\quad \mu_{bc}=\frac{m_bm_c}{m_b+m_c},
\eea
where $p_{XY}$ is the outgoing 4-momentum in the vertex with the particles $X$ and $Y$ in the loop. Finally, the constant $\lambda$ which governs the ratio of the $S$- and $D$-wave contributions reads
\be
\lambda=\frac{k_0^2}{\omega(k_0)}\sqrt{\frac{\Br(D_{s1}^+\to D^{*+}K^0)_{S\text{-wave}}}{\Br(D_{s1}^+\to D^{*+}K^0)_{D\text{-wave}}}},
\ee
where $k_0=149$~MeV is the final-state momentum in the decay $D_{s1}^+\to D^{*+}K^0$. Given sizeable uncertainties in the data on the $S$- and $D$-wave branchings of the $D_{s1}$ decay contained in the PDG tables \cite{Zyla:2020zbs} and assuming that these elastic decays fully saturate the width of the $D_{s1}$, we use the following representative 
values
\be
\Br(D_{s1}^+\to D^*K^+)_{S\text{-wave}}=0.61,\quad \Br(D_{s1}^+\to D^*K^+)_{D\text{-wave}}=0.39,
\ee
which give $\lambda\approx 53~\mbox{MeV}.$

In principle, the coupling constants introduced in the amplitudes \eqref{eq:A1}-\eqref{eq:A2} can be expressed through the couplings from the Lagrangians (\ref{L_YTH})-(\ref{L_THKD}) and (\ref{L_YHHK}). We, however, refrain from using these explicit relations and introduce coupling constants in the above mentioned amplitudes to be fitted to the data directly for two reasons. First, to simplify notations, all constant factors can be absorbed in the overall normalisation of the production rate which is a fitting parameter. Second, while the bare coupling constants connecting both fields $D_{s1}$ and $D_{s2}$ with $Y(4660)$ can indeed be related with $g_{\rm JT H}$ and $g_{\rm THK}^{(S,D)}$ from Eqs.~(\ref{L_YTH})-(\ref{L_THKD}), the corresponding dressed couplings may be different. Therefore, instead of one common constant we introduce two independent couplings $g_{D_{s1}}$ and $g_{D_{s2}}$ for the amplitudes proceeding through the $D_{s1}$ and $D_{s2}$, respectively.  
Then, although the strengths of the production mechanisms via the $D_{s1}$, $D_{s2}$ and point-like source are fixed independently, 
 the Lagrangian densities from Eqs.~(\ref{L_YTH})-(\ref{L_THKD}) and (\ref{L_YHHK}) are used 
to provide the relevant momentum dependence of the vertices and the HQSS-based relations between the individual terms in the square brackets in Eqs.~\eqref{eq:A1}-\eqref{eq:A2}.

We note also that, while the triangle loop is finite, the scalar loop is divergent and regularised by a sharp cutoff $\Lambda$ with the regulator dependence 
largely absorbed in the bare production constant $g\equiv g(\Lambda)$. 

\subsection{Invariant mass distribution of the $\bar{D}_s D^*/ \bar{D}_s^* D$ meson pairs}
\label{Sec:rate}

Given the energy range measured by BES~III, it is natural to assume that the reaction $e^+e^-\to K^+(D^-_sD^{*0} + D^{*-}_s D^0)$ proceeds dominantly through the excitation of the $Y(4660)$ resonance in the intermediate state. Then, with the amplitudes defined in the previous section, the $\bar{D}_{s}^*D^0 + \bar{D}_{s}D^{*0}$ the differential cross section reads
\be
\frac{d \sigma}{dm_{23}}={\cal N} 
\left|\frac{2m_{Y}}{s-m_Y^2+im_Y\Gamma_Y}\right|^2 \sum_{\alpha=1}^2\frac{ k\, q^{(23)}[\alpha]}{s}
\int_{-1}^1\frac{dz}{2} \left[k^4\left|M_D[\alpha]\right|^2+\left|M_S[\alpha]\right|^2\right],
\label{eq:rate}
\ee
where $M_Y=4.633$ GeV, $\Gamma_Y=64$ MeV \cite{Zyla:2020zbs}, and ${\cal N}$ is the overall normalisation factor which subsumes the unknown coupling $Y(4660) \to e^+e^-$ as well as all other irrelevant constants. The Mandelstam invariant $t$ in the centre of mass of the final $D$ mesons (centre of mass of the system (23)) reads
\be
t[\alpha]=m_K^2+m_3^2+2E_1^{(23)}E_3^{(23)}[\alpha]-2p_1^{(23)}q^{(23)}[\alpha]z,\quad z=\cos\theta,
\label{eq:t}
\ee
where $\theta$ is the helicity angle; the momenta and energies of the
kaon ($p_1^{(23)}[\alpha]$ and $ E_1^{(23)}[\alpha]$) and $D$ mesons ($q^{(23)}[\alpha]$ and $ E_3^{(23)}[\alpha]$) are given by
\bea\label{eq:kinem23}
&&p_1^{(23)}=\frac{\sqrt{s}}{m_{23}} k,\quad E_1^{(23)}=\sqrt{m_K^2+{(p_1^{(23)})}^2} \nonumber\\[-2mm]
\\[-2mm]
&&q^{(23)}[\alpha]=\frac{\lambda^{1/2}(m_{23}^2,m_2^2,m_3^2)}{2m_{23}},\quad E_3^{(23)}[\alpha]=\sqrt{m_3^2+{\left(q^{(23)}[\alpha]\right)}^2},\nonumber 
\eea
where, for channel 1 the masses entering the amplitudes $M_S[1]$ and $M_D[1]$ as well as the kinematical variables in Eqs.~\eqref{eq:t}-\eqref{eq:kinem23} are $m_2=m_{\bar{D}_s}$ and $m_3= m_{D^{*0}}$ while for channel 2 -- 
$m_2=m_{{\bar{D}_s}^*}$ and $m_3=m_{D^0}$, and for the masses we use the values given by the PDG review 
\cite{Zyla:2020zbs}.

The function minimised in the fit (${\cal L}$ is the likelihood) is defined in a standard way \cite{Zyla:2020zbs},
\be
-2\log{\cal L}=2\sum_i\left(\mu_i-n_i+n_i\log\frac{n_i}{\mu_i}\right),
\label{Lh}
\ee
where the sum runs over all data points in the analysed invariant mass distribution for all values of $\sqrt{s}$ used in the BES~III measurements; 
$n_i$ is the number of events and $ \mu_i$ is the value of the theoretical signal function from Eq.~\eqref{eq:rate} at the $i$-th data point in $m_{23}$,
corrected for the detector efficiency ($\bar{\epsilon}$), integrated luminosity (${\cal L}_{\rm int}$), and correction factor ($f_{\rm corr}$) taken directly from Ref.~\cite{BESIII:2020qkh},
\be
\frac{dN}{dm_{23}} =\frac{d \sigma}{dm_{23}} { \bar \epsilon \, {\cal L}_{\rm int} \, f_{\rm corr}}.
\ee
The combinatorial background from Ref.~\cite{BESIII:2020qkh} is also added incoherently.

The coupling constant $g_{D_{s2}}$ is set to unity since it can be absorbed by ${\cal N}$. Therefore, we have 5 parameters to be fitted to the invariant mass distributions provided by BES~III, namely, two parameters in the elastic $D$-meson potentials, $\mathcal{C}_d$ and $\mathcal{C}_f $, two ratios of the coupling constants for the production through the $Y(4660)$, $g/g_{D_{s2}}$ and $g_{D_{s1}}/g_{D_{s2}}$, and the overall normalisation factor ${\cal N}$. 

\section{Results and Discussions}

\begin{table}[t!]
\begin{center}
\begin{tabular}{|l|cccccc|}
\hline
Fit & $\mathcal{C}_{d}$, fm$^2$ & $\mathcal{C}_{f}$, fm$^2$ & $g_{D_{s1}}/g_{D_{s2}}$ & $g/g_{D_{s2}}$ & ${\cal N}$, $10^{-2} \frac{\rm pb}{\rm GeV}$
& $-2\log{\cal L}$\\
\hline
fit 1 & $-0.51\pm0.02$ & $ 0.18\pm 0.02$ & $0.26\pm0.02$ & $-2.47\pm 0.30$ & $0.46\pm 0.05$ & $138$ \\
fit 1$^\prime$ & $-0.23\pm 0.05$ & $ -0.1\pm 0.05$ & $0.37\pm 0.03$ & $-2.9\pm 0.6 $ & $0.35\pm 0.04$ & $144$ \\
fit 2 & $0.48$ & $-1.02\pm 0.01$ & $-0.44\pm 0.03$ & $ -6.3\pm 2.4$ & $0.28\pm 0.03$ & $146$\\
\hline
\end{tabular}
\end{center}
\caption{
The parameters of the best fits. The quality of each fit can be assessed through the value quoted in the last column. Fit 1 corresponds the global minimum while fits 1$^\prime$ and 2 are selected local minima.}
\label{tab:par}
\end{table}

\begin{figure}[t!]
\includegraphics[width=\textwidth]{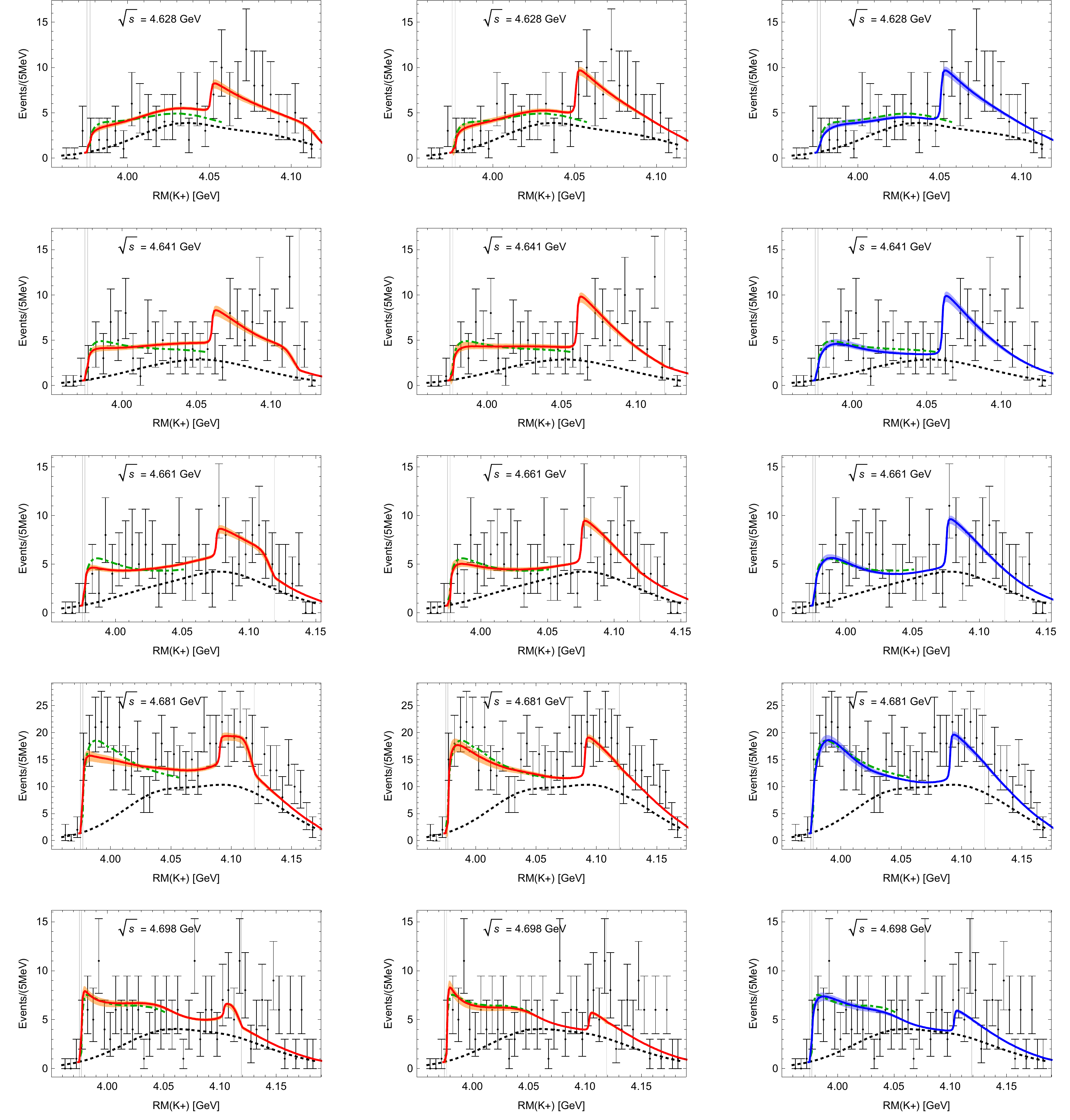} 
\caption{The line shapes corresponding to the best fits for the reaction $e^+e^- \to K^+(D_s^-D^{*0}+{D}_s^{*-}D^0)$ by BES~III \cite{BESIII:2020qkh} as functions of $\mbox{RM}(K^+)= \sqrt{p_{e+}^2+p_{e-}^2-p_{K^+}^2}$.
The results of fits 1 and 1$^\prime$ are shown as the red bands in the left and middle panels, respectively; the result of fit 2 is shown by the blue band in the right panel. The black dotted line stands for the combinatorial background from Ref.~\cite{BESIII:2020qkh}, while the thin gray vertical lines indicate the $D_s^-D^{*0}$, ${D}_s^{*-}D^0$, and $D^{*-}_sD^{*0} $ thresholds, in the order of the increasing energy. 
The green dot-dashed line shows the result of Ref.~\cite{Yang:2020nrt}. \label{Line_shape}}
\end{figure}

\begin{table}[tb]
\begin{adjustbox}{width=1.0\columnwidth}
\begin{tabular}{|l |cc | c c | c c| }
\hline 
\hline
$J^{P(C)}$ $\,$ State & \multicolumn{2}{c|}{Threshold, MeV} & RS & Poles fit $1$ & RS & Poles fit $1^\prime$
\\
\hline 
\hline
$1^{+\phantom{-}}$ $\,$ $Z_{cs}(3982)$ & $\bar{D}_s D^*/ \bar{D}_s^* D$ & $3975.2/3977.0$ & $(+++)$ & $3944\pm11$ & $(--+)$ & $3935_{-30}^{+5}$
\\
\hline 
$1^{+\phantom{-}}$ $\,$ $Z_{cs}(3982)$ & $\bar{D}_s D^*/ \bar{D}_s^* D$ & $3975.2/3977.0$ & $(--+)$ & $3971\pm2$ & $(--+)$ & $3972\pm2$
\\
\hline
$1^{+\phantom{-}}$ $\,$ $Z_{cs}^{\prime}$ & $\bar{D}_s^* D^*$ & $4119.1$ & $(--+)$ & $4115\pm2 -(9\pm2)i$ &$(++-)$& $ 4087_{-10}^{+10}+ 0_{-0}^{+47}i$
\\
\hline
\hline
$1^{+-}$ $\,$ $Z_c(3900)$ & $(D\bar D^*,-)$ & $3871.7$ & $(++)$ & $3842\pm10$ & $(-+)$ & $3832_{-38}^{+25}$ 
\\
\hline
$1^{+-}$ $\,$ $Z_c(4020)$ & $\bar{D}^* D^*$ & $4013.7$ & $(-+)$ & $4010\pm2 - (9\pm2) i$ & $(+-)$ & $3975_{-10}^{+15} + 0^{+45}_{-0}i $
\\
\hline
\hline
$1^{++}$ $\,$ $W_{c1}$ & $(D\bar D^*,+)$ & $3871.7$ & $(-)$ & $3864\pm2$ & $(-)$ & $3866\pm2$
\\
\hline
\hline
 $2^{++}$ $\,$ $W_{c2}$ & $\bar{D}^* D^*$ & $4013.7$ & $(-)$ & $4009\pm2$ & $(-)$ & $4011\pm2$
\\
\hline
\hline
\end{tabular}
\end{adjustbox}
\caption{Pole positions extracted from fits 1 and 1$^\prime$. The table shows the names of the states, their dominant decay channels and the corresponding thresholds, the Riemann sheet (RS) where they lie and the pole positions for the fits 1 and 1$^\prime$ (in MeV), in order. For each Riemann sheet, the signs of the imaginary parts of momenta in all coupled channels are quoted in parentheses.}
\label{tab:poles}
\end{table}

\begin{table}[tb]
\begin{center}
\begin{tabular}{|l |cc | c c | c c| }
\hline 
\hline
$J^{P(C)}$ $\,$ State & \multicolumn{2}{c|}{Threshold, MeV} & RS & Poles fit 2 
\\
\hline 
$1^{+\phantom{-}}$ $\,$ $Z_{cs}(3982)$ & $\bar{D}_s D^*/\bar{D}_s^* D$ & $3975.2/3977.0$ & $(+++)$ & $3954\pm2$ 
\\
\hline 
$1^{+\phantom{-}}$ $\,$ $Z_{cs}(3982)$ & $\bar{D}_s D^*/\bar{D}_s^* D$ & $3975.2/3977.0$ & $(--+)$ & $3958\pm7-(47\pm17) i$ 
\\
\hline
$1^{+\phantom{-}}$ $\,$ $Z_{cs}^{\prime}$ & $\bar{D}_s^*D^*$ & $4119.1$ & & --- 
\\
\hline
\hline
$1^{+-}$ $\,$ $Z_c(3900)$ & $(D\bar D^*,-)$ & $3871.7$ & $(-+)$ & $3863\pm7 -(59\pm13) i$
\\
\hline
$1^{+-}$ $\,$ $Z_c(4020)$ & $\bar{D}^* D^*$ & $4013.7$ & & --- 
\\
\hline
\hline
$1^{++}$ $\,$ $W_{c1}$ & $(D\bar D^*,+)$ & $3871.7$ & $(+)$ & $3852\pm2$
\\
\hline
\hline
 $2^{++}$ $\,$ $W_{c2}$ & $\bar{D}^* D^*$ & $4013.7$ & $(+)$ & $3990\pm2$ 
\\
\hline
\hline
\end{tabular}
\end{center}
\caption{Same as Table~\ref{tab:poles} but for fit 2. }
\label{tab:poles2}
\end{table}

There are two classes of solutions which provide the best overall fits to the BES~III data, which are shown in Fig.~\ref{Line_shape}. The parameters of the fits are listed in Table~\ref{tab:par}. In what follows, we refer to them as fit 1 and fit 2, respectively. Although they yield almost equally good description of the data, the corresponding physical pictures and the predictions from HQSS are very different, as discussed below. 
The poles of the amplitude for all the fits are given in Tables~\ref{tab:poles} and \ref{tab:poles2}.

We start from the common features of the both types of fits: 
\begin{itemize}
\item In both cases, there are poles near the $\bar{D}_s D^*/ \bar{D}_s^* D$ threshold which reveal themselves in the enhanced production rate right above this threshold. 
\item The rate near the $\bar{D}_s D^*/ \bar{D}_s^* D$ threshold is further enhanced by the triangle singularity from the $D_{s2} D_s^{*-} D^0$ loop.
\item All line shapes demonstrate a kind of plateau in the middle of the energy interval followed by a significant enhancement around 4.05-4.11 GeV, as required by the data at various production energies. This enhancement is driven by the tree-level $D_{s1}$ mechanism (see the  plot in the right panel of Fig.~\ref{fig:tree}). 
\end{itemize}

Now we come to the differences between fits 1 and 2. In Fig.~\ref{Line_shape} we present the results for fit 1 (left column). This solution demonstrates properties typical for a molecular picture. Specifically, in this case, the contact interaction potential in Eq.~\eqref{v3x3} is dominated by the elastic potential ${\cal C}_d$ (see Table~\ref{tab:par} for the values of the parameters) which equally contributes to all the diagonal transitions including the $\bar{D}_s^* D^{*0}\to \bar{D}_s^* D^{*0}$ one. Therefore, for this type of solution, if there is a molecular state near the $\bar{D}_s D^*/ \bar{D}_s^* D$ threshold associated with the $Z_{cs}(3982)$, there should also be a spin-partner state near the $\bar{D}_s^* D^{*0}$ threshold. 
The possibility to observe a structure corresponding to such a state in the $\bar{D}_s D^*/ \bar{D}_s^* D$ line shape and its actual position depends on the strength of the coupled-channel transition potential ${\cal C}_f$. If it is consistent with zero, no signature of the upper-threshold 
state will be seen in the lower channels. An example of such a scenario is the $Z_b(10650)$ as a $B^*\bar B^*$ molecule which does not show any clear peaking structures in the $B\bar B^*$ spectrum \cite{Voloshin:2016cgm,Wang:2018jlv} measured by Belle. For fit 1, the coupled-channel effects do generate a visible structure in the line shapes near the $\bar{D}_s^* D^{*0}$ threshold. Meanwhile, although some data points may indeed be enhanced in the relevant energy range, it is not possible to make a robust conclusion on the existence of such a structure in the BES~III data at least for the experimental accuracy available now. The results of fit 1 for the poles of the $Z_{cs}(3982)$ and its spin partners in the $\bar{D}_s^* D^{*0}$ and 
$\bar{D}^{(*)} D^{*0}$ channels are given in Table~\ref{tab:poles}. Due to the coupled-channel effects, there are two relevant poles\footnote{The poles on more remote Riemann 
sheets, which do not have an impact on the line shapes, are not shown.} near the $D_s^- D^{*0}$ threshold in this fit: 
(i) a bound state at $31$~MeV below the threshold, and
(ii) a virtual state at 4 MeV below the threshold.
The amplitude in the spin-partner channel with $J^{PC}=1^{+-}$, where the $Z_c(3900)$ resides, possesses a similar bound state with a comparable binding energy. Furthermore, in this fit, the $Z_{cs}^\prime$ and $Z_{c}(4020)$ are shallow quasi-bound states in the $\bar{D}_s^* D^*$ and $\bar{D}^* D^*$ channels, respectively, with the imaginary parts of the pole locations dominated by their coupling to the low-lying elastic (open-charm) channels. Meanwhile, the $J^{PC} = 1^{++}$ and $2^{++}$ spin-partner states, named here as $W_{c1}$ and $W_{c2}$, appear to be virtual states. 

Note that while fit 1 corresponds to the global
minimum of the function $-2\log{\cal L}$ defined in Eq.~(\ref{Lh}),
there exist also multiple local minima which yield a similar
description of the data and which can not be disentangled unambiguously
based on the current data only. As an example, we present fit 1$^\prime$ in Table~\ref{tab:par} (middle column in Fig.~\ref{Line_shape}). In this alternative fit of type 1, the coupled-channel transition potential ${\cal C}_f$ almost vanishes, so that the entire family of the fits of the first type can be characterised by the strong inequality
\be
|{\cal C}_f|\ll |{\cal C}_d|.
\label{fit1def}
\ee
The situation with the poles in fit 1$^\prime$ is in general similar to that
in fit 1, apart from the fact that now all the poles appear to be
virtual states with respect to the corresponding nearby thresholds
(see Table~\ref{tab:poles}). Therefore, while from the current data
alone it appears impossible to conclude whether the $Z_{cs}(3982)$ is
a bound or virtual state, if the scenario as outlined above for fits
of type 1 is realised, the $Z_{cs}(3982)$ is a spin partner of the $Z_c(3900)$ and $Z_c(4020)$ states. 

The results for the global-minimum fit 1 and especially fit 1$^\prime$ are consistent with those reported in Ref.~\cite{Yang:2020nrt} --- the corresponding line shapes are shown in Fig.~\ref{Line_shape} with the green dashed line which terminates at approximately 4.06 GeV (see the results for the cutoff 1 GeV in Figs. 3 and 4 of this Ref.). The lack of the $D_{s1}$ mechanism and coupled-channel effects does not allow one to employ the approach of Ref.~\cite{Yang:2020nrt} beyond this range. Since these effects are included in the current study, here the BES~III data are consistently described in the whole available energy interval, including the $\bar{D}^*_sD^*$ threshold --- see the text below for a discussion of the line shapes predicted for the $\bar{D}^*_sD^*$ channel and the information which can be extracted from them.

However, a quite different scenario allows one to describe the data almost equally well, although the corresponding fit (fit 2 in Table~\ref{tab:par}) has a slightly larger value of $-2\log{\cal L}$ than fits 1 and 1$^\prime$. The line shapes for fit 2 are shown in Fig.~\ref{Line_shape} (right column) and the poles of the amplitude are given in Table~\ref{tab:poles2}. Typical features of fit 2 include
\begin{itemize}
\item strong coupled-channel effects with the potentials in Eq.~\eqref{v3x3} satisfying the condition
\be
\Delta C = {\cal C}_d+{\cal C}_f< 0 \quad\mbox{with} \quad |\Delta C| \simeq |{\cal C}_d| \ll |{\cal C}_f| \quad\mbox{or} \quad |\Delta C|\ll |{\cal C}_d| \simeq |{\cal C}_f|,
\label{Eq:delta}
\ee
\item a strongly fine-tuned $\bar{D}_s D^*/\bar{D}_s^* D$ bound state. 
\end{itemize}

Like for the fits of type 1, we find a class of solutions of similar quality to that of fit 2 which correspond to very different individual values of ${\cal C}_d$ and ${\cal C}_f$ however with almost the same $\Delta C$ defined in \eqref{Eq:delta}. 

To better understand the two classes of solutions provided by the fits of types 1 and 2, consider, for simplicity, coinciding thresholds of the channels $\bar{D}_s D^*$ and $\bar{D}_s^* D$. Then, the poles of the multi-channel amplitude --- solution of the coupled-channel system \eqref{Eq:JPC} --- are defined by the equation
\be
\mbox{det}|I+G\cdot V|=\Bigl[({\cal C}_d+{\cal C}_f)J_0+1\Bigr]\Bigl[({\cal C}_dJ_0+1)({\cal C}_dJ_0'+1)-{\cal C}_f^2J_0J_0'\Bigr]=0,
\label{det}
\ee
where $I$ is the unit matrix, the matrix of the potentials $V$ is given by Eq.~(\ref{v3x3}), and $G=\mbox{diag}(J_0,J_0,J_0')$ with $J_0=J_0(D_s,D^*)=J_0(D_s^*,D)$ and $J_0'=J_0(D_s^*,D^*)$.

For the solutions of the first type we set ${\cal C}_f=0$ (see Eq.~(\ref{fit1def})), so that all coupled-channel effects vanish, and Eq.~(\ref{det}) reduces to
\be
({\cal C}_d J_0+1)^2({\cal C}_d J_0'+1)=0.
\ee
The poles of the amplitude are generated by the zeros of the expressions in the first and second parentheses in the expression above. 
Clearly, the existence of a pole as solution of the equation ${\cal C}_d J_0+1=0$ entails the existence of a similar partner pole near the $\bar{D}_s^*D^*$ threshold
from the equation ${\cal C}_d J_0'+1=0$. 

For the solutions of the second type we express equation (\ref{det}) in terms of $\Delta C$ introduced in Eq.~\eqref{Eq:delta} to arrive at
\be
(\Delta C J_0+1)\biggl ({\cal C}_d (J_0+J_0' + 2\Delta C J_0 J_0')-\Delta C^2 J_0 J_0'+1\biggr)=0.
\label{Eq:T}
\ee
Suppose the expression in the first parentheses vanishes, namely, 
\be\label{eq:bsfit2}
\Delta C J_0+1=0, \quad J_0\approx\frac{\mu}{\pi^2}\left(\Lambda+\frac{i\pi}{2}\sqrt{2\mu E}\right) + {\cal O}\left(\frac{E}{\Lambda}\right),
\ee
where $\mu$ is the reduced mass of the $D_s$ and $ D^*$. The amplitude then provides a bound state solution in the physically relevant region near the $\bar{D}_s D^*$ threshold, with $E=-E_B=-\gamma^2/(2\mu)$ and the binding momentum
\be
\gamma\approx\frac{2\pi}{\mu} \left( \frac1{\Delta C} +\frac{\mu}{\pi^2}\Lambda\right).
\ee
In other words, this scenario corresponds to a bound state controlled by the small parameter $\Delta C$ which appears as a result of severe cancellations between strongly correlated contact potentials ${\cal C}_d$ and ${\cal C}_f$. For $\Delta C \approx -0.55$ fm$^2$ from fit 2 (see Table~\ref{tab:par}) one finds a bound state about 18 MeV below the $\bar{D}_s D^*/\bar{D}_s^* D$ threshold. This rough estimate agrees well with the exact result for the poles for fit 2 given in Table~\ref{tab:poles2}\footnote{Since the results of fit 2 are almost insensitive to the other linear combination, ${\cal C}_d-{\cal C}_f$, when estimating the uncertainties in Table~\ref{tab:poles2} and in Fig.~\ref{Line_shape} the magnitude of this difference is fixed to the value which corresponds to the parameters for fit 2 quoted in Table~\ref{tab:par}.}. When the two thresholds are separated by the physical mass difference and the function $J_0$ from Eq.~\eqref{Eq:J} is utilised, this bound state is just somewhat shifted away from the threshold. 
In addition, there is a resonance pole in this fit, as shown in Table~\ref{tab:poles2}, emerging as a result of strong coupled-channel effects\footnote{Because of the Schwarz reflection principle this pole always has a mirror counterpart with an opposite sign of the Im part, which is not shown in Table~\ref{tab:poles2}.}. 

If this second scenario is realised, it entails very different consequences for the spin partner states compared to the fits of type 1 discussed
above. In particular, since the interaction in the $\bar{D}_s^* D^*$ channel is controlled by the contact term ${\cal C}_d$ 
(see the expression for the potential in Eq.~\eqref{v3x3}, which is positive in fit 2 (see Table~\ref{tab:par}), no
spin-partner states exist in this channel. To see this directly from Eq.~\eqref{Eq:T}, we notice that in the regime of the bound state given by Eq.~\eqref{eq:bsfit2}, the expression in the second parenthesis in Eq.~\eqref{Eq:T} reduces, up to a non-vanishing prefactor, to ${\cal C}_d J_0+1$, which can not vanish as long as ${\cal C}_d$ is positive. The absence of the $\bar{D}_s^* D^*$ spin partners can be clearly seen from
the line shape depicted in Fig.~\ref{Line_shape} which does not show any signal of a state near the $\bar{D}_s^* D^*$ threshold in spite of
strong coupled-channel effects ($|{\cal C}_f| > |{\cal C}_d|$). Note that this case is completely different to the one shown in Fig.~\ref{Line_shape} for fit 1$^\prime$ in which $|{\cal C}_f| \ll |{\cal C}_d|$ and, even though the $\bar{D}_s^* D^*$ spin partner exists, it is not seen in the $\bar{D}_s D^*$ line shape because of almost vanishing coupled-channel-driven transitions. On the other hand, the enhancement near the $\bar{D}_s D^*/ \bar{D}_s^* D$ threshold in fit 2 is still reproduced quite well.

If $Z_{cs}$ is governed by a bound state pole, the spin partners in the $1^{++}$ and $2^{++}$ channels driven by the same linear combination of the contact terms $\Delta C$ (see Eq.~\eqref{vfull}) should exist and have roughly the same binding energy as that for the $Z_{cs}(3982)$. However, similarly to the $\bar{D}_s^* D^*$ case, no fine-tuned states in the coupled channel with $J^{PC}=1^{+-}$ exist near the $D^*\bar{D}^* $ threshold. Nevertheless, a resonance state above the $D{\bar D^*}$ threshold may be generated by the strong coupled-channel dynamics in full analogy with the resonance poles near the $\bar{D}_s D^*/ \bar{D}_s^* D$ threshold. This picture, therefore, does not preclude the $Z_c(3900)$ from being a spin partner of the $Z_{cs}(3982)$, if both are treated as resonances. However, in this case, the state $Z_c(4020)$ is not their spin-partner and should be of a different nature. 
 
We note that for fits 1 and 1$^\prime$ both the pointlike and resonance (through $D_{sJ}$) production mechanisms provide equally important contributions to the line shapes, with their relative importance depending on the energy. In particular, for the smallest $\sqrt{s}$ the pointlike production mechanism dominates at least at low invariant masses while for the largest two energies 
provided by BES~III, the resonance mechanism is much more important. This pattern is generally consistent with the results shown in Fig.~5 of Ref.~\cite{Yang:2020nrt}. 
Contrary to fit 1, the results of fit 2 for all energies are almost completely dominated by resonance production through the $D_{sJ}$. 
However, these results are in conflict with the findings of Ref.~\cite{Du:2020vwb}, where $S$-wave kaon production was found to be more important at the single energy $\sqrt{s}=4.68$ GeV. 

\begin{figure}[t!]
\begin{center}
\includegraphics[width=0.5\textwidth]{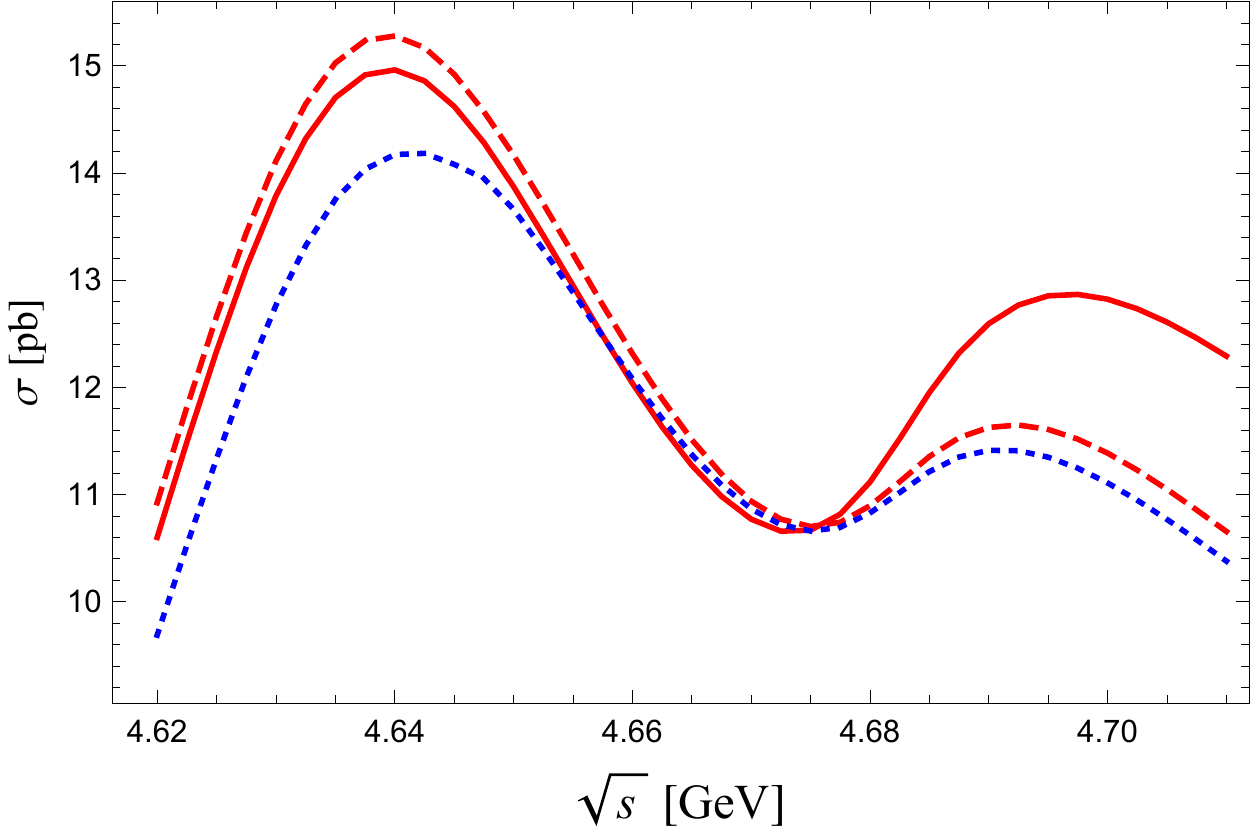}
\caption{Total cross section of the reaction $e^+e^-\to K^+(D_s^-D^{*0}+D_s^{*-}D^0)$ predicted for the reaction mechanisms shown in Fig.~\ref{fig:diag}.
The red solid, red dashed, and blue dashed lines correspond to the global fit 1, fit 1$^\prime$, and fit 2, respectively. 
\label{Fig:XS}}
\end{center}
\end{figure}

{ In Fig.~\ref{Fig:XS} we plot the total cross of the reaction $e^+e^-\to K^+(D_s^-D^{*0}+D_s^{*-}D^0)$ evaluated as given in Eq.~\eqref{eq:rate} 
for the mechanisms shown in Fig.~\ref{fig:diag}. This cross section exhibits two clear peaks near 4.64 and 4.69 GeV governed by the resonance 
mechanisms via the $D_{s1}$ and $D_{s2}$, that clearly illustrates their importance. In Ref.~\cite{BESIII:2020qkh}, the BES~III Collaboration listed the values of the product $\sigma^{\rm B} (e^+e^-\to K^+ Z_{cs}(3982)) \times {\rm BF}(Z_{cs}(3982)\to (D_s^-D^{*0}+D_s^{*-}D^0))$ of the Born cross section of the $Z_{cs}(3982)^-$ production and the sum of BFs of its decays into $D_s^-D^{*0}+D_s^{*-}D^0$ measured at several selected points in $\sqrt{s}$. These values appear to be significantly smaller than those we plot in Fig.~\ref{fig:diag}, though a peak near 4.69 GeV is also seen by BES~III. A direct comparison of these two results is, however, not possible since in Ref.~\cite{BESIII:2020qkh} all contributions to the cross section from the $D_{s1}$ and $D_{s2}$, treated as a background, were subtracted. Meanwhile, it  follows from the current analysis 
as well as  from the analysis of Ref.~\cite{Yang:2020nrt} that especially the $D_{s2}$ not only provides a background but also strongly enhances the production of the $Z_{cs}(3982)$.}

\begin{figure}[t!]
\begin{center}
\includegraphics[width=0.5\textwidth]{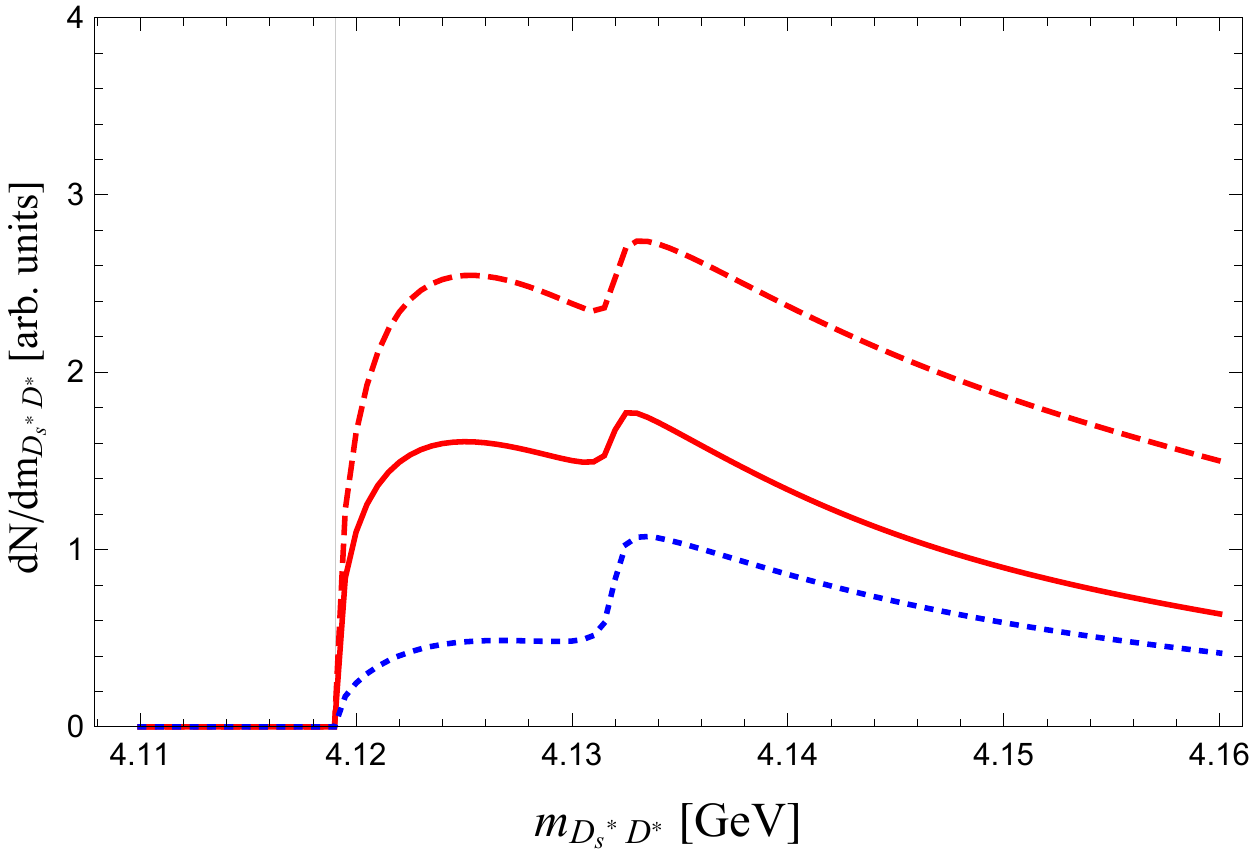} 
\caption{Invariant mass distributions of the $D^*$-meson pair 
in the reaction $e^+e^-\to K^+D^{*-}_sD^{*0} $ for $\sqrt{s}=4.681$ GeV predicted based on the best fits to the BES~III data in the $D_sD^*/D^*_sD$ channel. See the caption of Fig.~\ref{Fig:XS} for the colour scheme for the lines. 
The thin gray vertical line indicates the position of the $D^{*-}_sD^{*0} $ threshold. 
\label{Line_shape_DstDst}}
\end{center}
\end{figure}

Additional information that could be very useful to discriminate between different scenarios supported by the present BES~III data and impose additional constraints on the parameters of the contact interactions can be gained from the reaction $e^+e^-\to K^+D^{*-}_sD^{*0}$. To illustrate this point, in Fig.~\ref{Line_shape_DstDst}, we give the $\bar{D}^*_sD^*$ invariant mass distributions in the $J^{P}=1^+$ channel predicted for the parameters from Table~\ref{tab:par}. It is remarkable that the shape and the strength of this spectrum changes significantly from fit to fit. Indeed, although all three line shapes possess a structure near 4.133~GeV which comes from the tree-level diagram via $D_{s1}$, the curve for fit 2 (blue dotted line), which has no poles near the $\bar{D}^*_sD^*$ threshold, grows smoothly from the threshold while those for fits 1 and 1$^\prime$ demonstrate a significant enhancement in the near-threshold region that is a clear manifestation of an interplay of the near-threshold poles in the $\bar{D}^*_sD^*$ channel and the nearby triangle singularity. 

\section{Summary}

A coupled-channel analysis of the reaction $e^+e^- \to K^+(D_s^-D^{*0}+{D}_s^{*-}D^0)$ is presented for the whole energy range measured by BES~III. Two reaction mechanisms are involved: (a) production of the $\bar{D}_s^{(*)}D^{(*)}$-meson pairs from a point-like source and (b) excitation of the $D_{sJ} \bar{D}_s^{(*)}$ $(J=1,2)$ with the subsequent decay $D_{sJ}\to K D^{(*)}$. The final-state interaction within the $\bar{D}_s^{(*)}D^{(*)}$-meson pairs is included in the contact effective-field theory involving leading momentum-independent potentials between coupled channels ${D}_s^{*-}D^0$, $D_s^-D^{*0}$, and $D_s^{*-}D^{*0}$ while the influence of the $SU(3)$ meson exchanges on the results is to be clarified in future studies. 

The production process via the excited $D_{sJ}$ states is found to play an important role for the invariant mass spectra: The tree-level amplitudes via $D_{s2}$ and $D_{s1}$ provide visible enhancements in the spectra at low and relatively large invariant masses, respectively, which are consistent with the data. The triangle loop with the $D_{s2}D_s^*D$ intermediate state amplifies the pole contribution in the $\bar{D}_s^*D$ scattering amplitude which reveals itself in a pronounced enhancement of the production rate at low energies. This conclusion is consistent with the findings of Ref.~\cite{Yang:2020nrt}. The nearby singularities from the $D_{s2}D_s^*D^*$ and $D_{s1}D_s^*D^*$ triangle loops are found to be less important though 
they provide some support for the spin partner state near the $\bar D_s^*D^*$ threshold, when it exists. This suppression can be explained by the fact that such triangle loops are supplemented with the off-diagonal $T$-matrix elements which are suppressed in the fits possessing the poles near the $\bar D_s^*D^*$ threshold. In addition, the kaon momentum 
produced from the $D_{sJ}$ decay to $D^{(*)}K$ in a $D$-wave provides a relative suppression of the spectra at larger invariant masses for the two $D$-mesons, where the peaks from the $D_{s2}D_s^*D^*$ and $D_{s1}D_s^*D^*$ triangles are located. 

We find two classes of solutions which describe the data almost equally well, but
strongly differ in the underlying dynamics and, therefore, also in the predictions for the spin partner states. One class is characterised by a clear dominance of the diagonal transitions in the effective potential (fits 1 and 1$^\prime$). The $Z_{cs}(3982)$ in this case is identified as a molecular bound or virtual state below the ${D}_s^{*-}D^0$ threshold which (i) appears naturally as a $SU(3)$ partner of the $Z_c(3900)$ and $Z_c(4020)$ states and (ii) has a spin partner near the $D_s^{*-}D^{*0}$ threshold. The other class of solutions, which is only slightly less preferred by the maximum-likelihood fits, corresponds to a very strong coupled-channel dynamics with the off-diagonal transitions being larger or comparable with the diagonal ones. In this case, the $Z_{cs}(3982)$ can still be a spin partner of the $Z_c(3900)$ but only if both show up as resonance states with sizeable imaginary parts of the poles generated by the significant coupled-channel effects. While this is still consistent with the BES~III data on the $e^+e^-$ annihilation into the final state $K^+(D_s^-D^{*0}+D_s^{*-}D^0)$ it remains to be seen if such a pole is consistent with the line shapes measured for the $Z_c(3900)$. If this scenario is realised, the $Z_c(4020)$ can not be a spin partner of the $Z_c(3900)$ and $Z_{cs}(3982)$ and should have a different nature. Also, no spin partner of the $Z_{cs}(3982)$ near the $D_s^{*-}D^{*0}$ threshold exists in this case. In order to distinguish between the two solutions, we predict the form of the $J^{P}=1^+$ $\bar{D}_s^*D^*$ invariant mass distribution to be measured in the reaction $e^+e^-\to K^+D_s^{*-}D^{*0}$ and argue that the corresponding data should shed light on the nature of the $Z_{cs}(3982)$ and its spin partners. 

\section*{Acknowledgments}

The authors would like to thank F.-K. Guo and J. Nieves for reading the manuscript and valuable comments.
This work is supported in part by the National Natural Science Foundation of China (NSFC) and the Deutsche Forschungsgemeinschaft (DFG) 
through the funds provided to the Sino- German Collaborative Research Center TRR110 ``Symmetries and the Emergence of Structure in QCD'' (NSFC Grant No. 12070131001, DFG Project-ID 196253076) and by the German Ministry of Education and Research (BMBF) (Grant No.~05P21PCFP4). 
The work of A.N. is supported by the Ministry of Science and Education of the Russian Federation under grant 14.W03.31.0026.

\end{document}